\documentclass[[trackchanges,twocolumn]{aastex63}

\usepackage{subfigure}
\usepackage{soul}
\usepackage{amsmath}	
\usepackage{amssymb}	

\usepackage{xcolor}

\received{August 5, 2020}
\accepted{\today}
\submitjournal{AJ}

\shorttitle{ZTF source classification I}
\shortauthors{J. van Roestel et al}
\graphicspath{{./}{Figs/}}

\begin{document}

\title{The ZTF Source Classification Project: I. Methods and Infrastructure} 

\correspondingauthor{Jan van~Roestel}
\email{jvanroes@caltech.edu}

\author[0000-0002-2626-2872]{Jan van~Roestel}
\affiliation{Division of Physics, Mathematics and Astronomy, California Institute of Technology, Pasadena, CA 91125, USA}

\author[0000-0001-5060-8733]{Dmitry A. Duev}
\affiliation{Division of Physics, Mathematics and Astronomy, California Institute of Technology, Pasadena, CA 91125, USA}

\author[0000-0003-2242-0244]{Ashish A. Mahabal}
\affiliation{Division of Physics, Mathematics and Astronomy, California Institute of Technology, Pasadena, CA 91125, USA}

\author[0000-0002-8262-2924]{Michael W. Coughlin}
\affiliation{School of Physics and Astronomy, University of Minnesota, Minneapolis, Minnesota 55455, USA}

\author[0000-0001-7016-1692]{Przemek Mr{\'o}z}
\affiliation{Division of Physics, Mathematics and Astronomy, California Institute of Technology, Pasadena, CA 91125, USA}

\author[0000-0002-7226-836X]{Kevin Burdge}
\affiliation{Division of Physics, Mathematics and Astronomy, California Institute of Technology, Pasadena, CA 91125, USA}

\author{Andrew Drake}
\affiliation{Division of Physics, Mathematics and Astronomy, California Institute of Technology, Pasadena, CA 91125, USA}

\author[0000-0002-3168-0139]{Matthew J. Graham}
\affiliation{Division of Physics, Mathematics and Astronomy, California Institute of Technology, Pasadena, CA 91125, USA}

\author{Lynne Hillenbrand}
\affiliation{Division of Physics, Mathematics and Astronomy, California Institute of Technology, Pasadena, CA 91125, USA}

\author[0000-0001-8018-5348]{Eric C. Bellm}
\affiliation{DIRAC Institute, Department of Astronomy, University of Washington, 3910 15th Avenue NE, Seattle, WA 98195, USA}

\author{Alexandre Delacroix}
\affiliation{Caltech Optical Observatories, California Institute of Technology, Pasadena, CA  91125}

\author[0000-0002-4223-103X]{C.~Fremling}
\affiliation{Cahill Center for Astrophysics, 
California Institute of Technology, MC 249-17, 
1200 E California Boulevard, Pasadena, CA, 91125, USA}

\author[0000-0001-8205-2506]{V. Zach Golkhou}
\affiliation{DIRAC Institute, Department of Astronomy, University of Washington, 3910 15th Avenue NE, Seattle, WA 98195, USA} 

\author{David Hale}
\affiliation{Caltech Optical Observatories, California Institute of Technology, Pasadena, CA  91125}

\author[0000-0003-2451-5482]{Russ R. Laher}
\affiliation{IPAC, California Institute of Technology, 1200 E. California Blvd, Pasadena, CA 91125, USA}

\author[0000-0002-8532-9395]{Frank J. Masci}
\affiliation{IPAC, California Institute of Technology, 1200 E. California Blvd, Pasadena, CA 91125, USA}

\author[0000-0002-0387-370X]{Reed Riddle}
\affiliation{Caltech Optical Observatories, California Institute of Technology, Pasadena, CA  91125}

\author[0000-0002-6099-7565]{Philippe Rosnet}
\affiliation{Université Clermont Auvergne, CNRS/IN2P3, LPC, Clermont-Ferrand, France}

\author[0000-0001-7648-4142]{Ben Rusholme}
\affiliation{IPAC, California Institute of Technology, 1200 E. California Blvd, Pasadena, CA 91125, USA}

\author[0000-0001-7062-9726]{Roger Smith}
\affiliation{Caltech Optical Observatories, California Institute of Technology, Pasadena, CA  91125}

\author[0000-0001-6753-1488]{Maayane T. Soumagnac}
\affiliation{Lawrence Berkeley National Laboratory, 1 Cyclotron Road, Berkeley, CA 94720, USA}
\affiliation{Department of Particle Physics and Astrophysics, Weizmann Institute of Science, Rehovot 76100, Israel}

\author{Richard Walters}
\affiliation{Caltech Optical Observatories, California Institute of Technology, Pasadena, CA 91125}


\author{Thomas A. Prince}
\affiliation{Division of Physics, Mathematics and Astronomy, California Institute of Technology, Pasadena, CA 91125, USA}

\author[0000-0001-5390-8563]{S. R.~Kulkarni}
\affiliation{Cahill Center for Astrophysics, California Institute of Technology, MC 249-17,1200 E California Boulevard, Pasadena, CA, 91125, USA}




\begin{abstract}
The Zwicky Transient Facility (ZTF) has been observing the entire northern sky since the start of 2018 down to a magnitude of 20.5 ($5 \sigma$ for 30s exposure) in $g$, $r$, and $i$ filters. Over the course of two years, ZTF has obtained light curves of more than a billion sources, each with 50-1000 epochs per light curve in $g$ and $r$, and fewer in $i$. To be able to use the information contained in the light curves of variable sources for new scientific discoveries, an efficient and flexible framework is needed to classify them. In this paper, we introduce the methods and infrastructure which will be used to classify all ZTF light curves. Our approach aims to be flexible and modular and allows the use of a dynamical classification scheme and labels, continuously evolving training sets, and the use of different machine learning classifier types and architectures. With this setup, we are able to continuously update and improve the classification of ZTF light curves as new data becomes available, training samples are updated, and new classes need to be incorporated. 

\end{abstract}






\section{Introduction}
Astronomy, like many other branches of science, has been experiencing an explosive increase in data volumes, which are doubling roughly every two years. This revolution has driven a renaissance in many areas of astronomy, most notably in the time domain. At some level, all astronomical sources exhibit changes in their brightness with time, driven by a myriad of different phenomena. The study of source variability has benefited greatly from the data deluge providing insight into a broad range of astrophysical processes and phenomena.

The light curves of variable objects contain information about the nature of the objects and the physical processes that are responsible for the observed changes. Variable objects are a key tool in astrophysics and are the main science driver in many fields. While it is nearly impossible to list all their astrophysical applications, variable stars have been used as distance indicators \citep[e.g.,][]{pietrzyn2013,pietrzyn2019,riess2018}, tracers of the structure and kinematics of the Milky Way and nearby galaxies \citep[e.g.,][]{skowron2019,chen2019,ajd2016,ajd2017}, or tracers of the chemical evolution of galaxies \citep[e.g.,][]{genovali2015}. Studying stellar variability also helps us to understand the evolution and physics of stars themselves -- detailed modeling of eclipses has enabled precise measurements of masses and radii of all types of stars \citep[e.g.,][]{torres2010}, asteroseismology is being used to great effect to study the interior structure of stars \citep[e.g.,][]{aerts2019}, and the irregular variability of cataclysmic variables (CV), young stellar objects (YSO), and active galactic nuclei (AGN) offers insight into accretion physics on all scales \citep{scaringi2015}.

The necessary first step in enabling all these applications is to identify variable sources and classify them into known object types while simultaneously looking for new classes. 

The astronomical community has extensive experience in dealing with large samples of light curve data enabled by survey telescope automation and advances in both the camera technology and data processing and analysis techniques. Notable examples of large-scale surveys are: the All Sky Automated Survey \citep[ASAS;][]{pojmanski1997}, the All Sky Automated Survey for Supernovae \citep[ASAS-SN;][]{shappee2014}, the Asteroid Terrestrial-impact Last Alert System \citep[ATLAS;][]{tonry2018}, the Catalina Real-Time Transient Survey \citep[CRTS;][]{drake2014}, EROS \citep{tisserand2007}, Gaia \citep{gaia2016}, MACHO \citep{alcock2000}, the Northern Sky Variability Survey \citep[NSVS;][]{wozniak2004} the Optical Gravitational Lensing Experiment \citep[OGLE;][]{udalski2003,udalski2015}, Pan-STARRS1 \citep{PS1}, the VISTA Variables in the Via Lactea \citep[VVV;][]{minniti2010}. 

To deal with the massive amount of data involved, these projects usually employ machine learning (ML) techniques to detect and classify variable sources \citep[e.g.,][]{wozniak_ml2004,debosscher2007,kim2011,kim2014,palaversa2013,masci2014,armstrong2016,heinze2018,holl2018,jayasinghe2019,jayasinghe2020}. However, a more traditional approach -- with the light curves vetted by a human expert -- also proves to be successful \citep[e.g.,][]{drake2014,soszynski2014,soszynski2015,soszynski2016,soszynski2016b,udalski2018}. To date, over a million variable stars have been detected and classified, the majority of which were found by OGLE \citep{soszynski2018}.

Astronomical light curve data are typically sparsely and unevenly sampled, incomplete, heteroskedastic, and come with a lot of different biases. It is challenging to apply standard time series processing and analysis techniques developed in other areas to such data.

A common approach to classification is to first compute a set of summary statistics (features), such as the mean or median flux, interquartile range (iqr), von Neumann ratio, period(s)/amplitude(s), etc. These features encode the light curves (with different cadences and number of epochs) as a vector of finite length which allows for direct comparison of objects. \citet{debosscher2007,bloom2012,nun2015,kim2016} use those to classify the objects. This task can be done by humans (often by inspecting only two features at a time); but the scale of the problem essentially forces one to use machine learning methods.


At the forefront of the revolution in time-domain astronomy, the Zwicky Transient Facility (ZTF) project uses the 48-inch (1.2 meter) Samuel Oschin Schmidt telescope at Palomar Observatory in Southern California to observe the sky every night. Science observations began on March 17th, 2018 \citep{Graham2019, Bellm2019}. The median magnitude limit is 20.5 in the $r$ band for a nominal 30-second exposure time ($5\sigma$ detection). ZTF has been performing frequent accurate measurements of more than a billion astronomical objects observable from Palomar Observatory (declination$>-28^{\circ}$).

ZTF light curves of variable stars have already been used to make exciting discoveries by using targeted searches. Searching for very short period variability, \citet{BuCo2019} discovered one of the shortest period binary systems known with a period of just 7 minutes. ZTF light curves have also been used to discovered new types of variable stars: \citet{kupfer2020a} discovered a new type of compact objects binary, and \citet{kupfer2020} found a new type of pulsating star. \citet{Vanderbosch2019} discovered a white dwarf with exocomets, the second of such a system.
ZTF has also been used to discover large number of outbursting or flaring objects: cataclysmic variables and microlensing events \citet{szkody2020,Mroz2020}. 

In this paper, we present the framework designed by the ZTF project to identify and classify variable objects in all ZTF data. Section 2 describes the ZTF light curve data including pre-processing and feature extraction. In Section 3, we introduce the classification scheme adopted for ZTF and describe the ML algorithms used therein. In Section 4, we present the active learning approach to labeled data set assembly and classifier training. The performance of the resulting classifiers are discussed in Section 5. Finally, in Section 6 we discuss this initial study and outline our future work plans.


\section{ZTF light curves and pre-processing} 

\begin{figure*}
    \centering
    \includegraphics{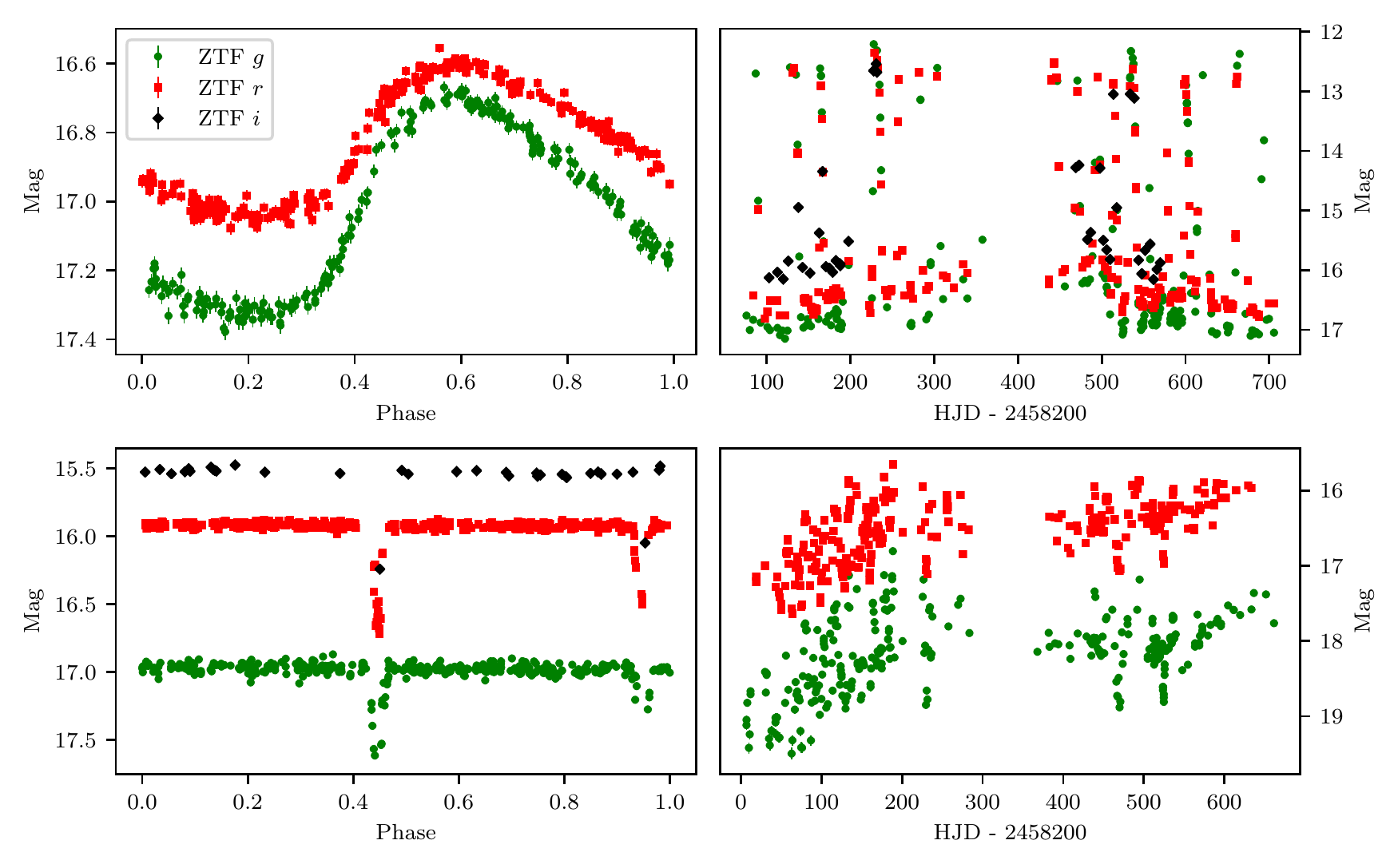}
    \caption{Example light curves from ZTF. The left column shows two periodic variables, an RR Lyrae and an eclipsing binary. The right column shows an outbursting cataclysmic variable and an irregular variable young stellar object. Note that we classify the individual light curves for objects.}
    \label{fig:exampleLCs}
\end{figure*}

\subsection{ZTF light curves}
The ZTF camera uses 16 separate $6k \times 6k$ CCD detectors and has a total field of view of 47 square degrees with a pixel size of 1.01$\arcsec$ \citep{Bellm2019}. ZTF pointings are organized in two grids with rows of equal declination to cover the entire northern sky ranging from declination $-28^{\circ}$ to the Northern celestial pole. The primary grid uses 637 pointings of slightly overlapping ``fields'', covering 88\% of the observable sky. The remaining area falls in the gaps between CCD detectors. To cover this missing area, a secondary grid (897 pointings), offset in both right ascension and declination from the primary grid, is used.

Several surveys are carried out by ZTF which use different filters, cadences, sky areas, and exposure times. The main public survey (40\% of the time) is an all-sky survey in $g$ and $r$ with a cadence of 3 days \citep{Bellm2019}. Smaller, dedicated surveys are carried out by the ZTF partnership (40\%) and time available to Caltech (20\%). The largest survey is the supernova survey with six observations per night of $\approx$3000 square degrees,  a survey of the TESS footprint, and other smaller surveys. ZTF also carried out deep-drilling observations of the Galactic Plane, where one field was typically observed continuously for 1.5\,hrs. Most of these surveys use $g$- and $r$-bands (deep-drilling is only done in $r$-band), but a small fraction of the observations ( $\sim 1.6$\%), are in $i$-band. Most of the surveys focus on observing fields from the primary grid. 

At the time of writing, the median (min, max) number of epochs for primary grid fields for all surveys combined are 184 (26, 1079) in $g$, 338 (23, 1263) in $r$, and 23 (1, 165) in $i$. This broad range in the number of epochs per field is partially due to observability (lower declination fields tend be have fewer epochs), but mostly because of the smaller surveys which tend to accumulate many epochs for small sets of fields. 
The median number of epochs per field in the secondary grid is much lower (median $<50$ for all filters). The low number of epochs makes classification challenging, and we did not include them in this study but will do so in future work.

All ZTF images are processed and data products are automatically generated. This includes light curves of all persistent sources in the science images, which are the main data product for this work. Here, we summarise the process, for a full description, see \citet{Masci2019}.

First, reference source lists are generated by running a source finding algorithm on reference images. Reference images are constructed by combining at least 15 images of good quality. When new science images are available, \texttt{SExtractor} \citep{bertin1996} is applied and sources within 1.5$\arcsec$ of a reference source are linked to that reference source to construct a light curve. Each filter, and each CCD-quadrant\footnote{each CCD has four readout channels} per ZTF pointing is processed completely separately from all other data. This means that a single astrophysical object will have multiple ZTF light curves for each filter, and if they occur in multiple ZTF fields (which can occur in the overlap between fields or the primary and secondary grid), will have multiple light curves even for the same filter.

In this work, we will use the individual light curves as the basis for our classification, see for example Fig. \ref{fig:exampleLCs}. While combining light curves potentially allows for better classification, we choose to classify the individual light curve instead of combining them. We do this for several reasons. First, the large field of view of ZTF makes perfect absolute calibration of light curves difficult. Combining light curves with small but significant calibration differences will introduce spurious variability. Second, image artifacts (ghosts on the CCD, bad pixels, etc) are position-dependent and typically only affect one light curve of an object. Keeping the light curve separate allows the objects to be classified using the unaffected light curves. An additional motivation to not combine light curves is that not all objects have light curves in the different band-passes, especially for faint and/or red objects. Classifying only single-band light curve allows for a more uniform classification. Note that we do inspect all light curves simultaneously when labeling light curves, and light curves of the same object share the same label.

Besides the light curves, images are also processed by a difference image pipeline, and any source more significant than $5\sigma$ on the difference images (positive or negative), is reported as an alert \citep{masci2018}. While this pipeline is mainly designed to study transients and moving objects, variable sources also generate alerts. While, in principle, the alerts do not contain new information, the separate pipeline allows for a consistency check which is useful to identify image and processing artifacts. We, therefore, include some information from the alert pipeline in our analysis.

\subsection{Light curve pre-processing}

Two main approaches to the light curve classification problem have been employed by the community, differing by what is fed into a machine learning system: either pre-computed features  \citep[e.g.][]{blomme2010,richards2012} or the light curves directly \citep{Naul2017, MuNa2019, jamal2020}.

The first approach inevitably causes certain information loss, even though the computed features provide a powerful and standardized insight into the raw data. The choice of such features referred to as ``feature engineering'' in the ML world, is a highly non-trivial problem on its own.

The recent success of techniques that use artificial many-layer neural networks \citep[deep learning, DL;][]{McCulloch1943},
 is in big part attributed to the ability of such systems to discover and extract relevant features directly from the data. DL systems frequently outperform more traditional approaches; however, it is challenging to apply those techniques to astronomical data due to the intrinsic characteristics of the data discussed above \citep{Naul2017}.

In this work, we employ a hybrid approach to retain the advantages of both methods. We rely on the light curve features while simultaneously striving to preserve more information contained in the time series by using a two-dimensional second-order mapping of the light curves based on the changes in magnitude (\textit{dm}) over the available time-differences (\textit{dt}) \citep{mahabal2017}.


\subsubsection{Calculation of light curve features}
For each light curve, we calculate a number of simple statistics, determine the best period and significance using a period-finding algorithm, and evaluate features that are the result of fitting the phase-folded light curve with a multi-harmonic sinusoid. Here, we briefly summarise the procedure; for full details, please see \citet{CoBu2020b}. 
First, we remove any light curve epochs which are flagged as taken in bad conditions, which is about 6\% of all data. We also skip any object within an empirically chosen radius 13$^{\prime\prime}$ of ``bright stars,'' taken to be stars in Gaia \citep{Gaia2018} brighter than 13th magnitude or any object in the Yale Bright Star Catalog \citep{HoJa1991}.

Early experiments showed that the feature values are strongly affected by the presence of deep-drilling data in the light curves. E.g. deep-drilling observations of long-period variables at one particular phase of the light curve significantly skew many of the light curve statistics. Because this would severely limit the use of the features, we decided to mask deep-drilling data when calculating the features.
As shown by \citet{pashchenko2018}, many of the commonly used light curve features are strongly correlated. We therefore only calculate a small set of features. We did add a few redundant features to the set suggested by \citet{pashchenko2018} (e.g., $\chi^2$, inter-percentile ranges). These features allowed us to better assess the quality of the light curves.

\subsubsection{Period finding and Fourier features}
The period finding strategy relies on a hierarchical technique, where two fast algorithms, conditional entropy (CE; \citealt{GrDr2013}) and Lomb-Scargle (LS, \citealt{Lo1976}; \citealt{Sc1982}), are used to identify high-significance, candidate periods, which are then passed to a slower, more comprehensive algorithm, multi-harmonic analysis of variance (AOV, \citealt{ScCz1998}).
Our fast algorithms are implemented on Graphics Processing Units (GPU) in CUDA, the specific implementation of CE can be found in \citealt{KaCo2020}\footnote{https://github.com/mikekatz04/gce} and the LS implementation can be found here\footnote{https://github.com/johnh2o2/cuvarbase}. A CPU-based AOV is then applied to the top 50 frequencies identified by each of the algorithms to identify the best period. We again masked any deep-drilling data as it strongly affects the period-finding performance.

Once the best period has been identified, we fit the light curve with a simple model that combines an offset and slope with a series of sinusoids using that period. The model is described by:
\begin{equation}
M(t) = st+c + \sum_{n=1}^{n=5} a_n \sin(n \frac{2\pi t}{P})+b_n \cos(n \frac{2\pi t}{P})
\end{equation}
The parameters $a_n$ and $b_a$ are converted to amplitudes and phases, and the amplitudes and phases of the harmonics normalized to the amplitude of the first harmonic. To determine the goodness-of-fit of this model, we use the Bayesian Information Criterion (BIC) value \citep{schwarz1978}. The number of harmonics used is determined by the lowest BIC value. 

\subsubsection{Magnitude-time histograms -- `\textit{dmdt}'}
As additional input for the deep-learning-based classifiers, we calculate a 2D histogram from all pairs of magnitude and time difference (\textit{dm} and \textit{dt}, respectively). This method encodes the one-dimensional light curves of various lengths into a two-dimensional array of fixed dimensions (an image), which is much easier for a classifier to interpret \citep{mahabal2017}. We use 26 approximately logarithmic spaced time bins and 26 magnitude bins, approximately logarithmic in both positive and negative magnitude differences. We did include deep-drilling data into the calculation of the histograms as the high cadence data would fall mostly in the low-\textit{dt} bins that are not populated by the rest of the data points.

\subsection{External data}
In addition to the data based purely on the ZTF light curves, some of our classifiers use data extracted from external catalogs. We spatially cross-matched all of the ZTF objects with the AllWISE \citep{WrEi2010}, Gaia DR2 \citep{GaiaDR22018}, and Pan-STARRS1 DR1 \citep{PS1}  catalogs using a match radius of 2$\arcsec$ and extracted the following data (and a catalog ID) for the closest corresponding object within that radius:
\begin{itemize}
    \item Gaia DR2: the $G$, $BP$ and $RP$ magnitudes, the parallax and proper motion with their associated uncertainties
    \item Pan-STARRS1 DR1: the $grizy$ magnitudes with their uncertainties
    \item AllWISE: the $W1$, $W2$, $W3$, and $W4$ magnitudes and their uncertainties.
\end{itemize}

\subsection{Data storage and access}

Efficient data storage and access, given the data set size, represent a substantial problem. We solved it by employing \texttt{Kowalski}\footnote{https://github.com/dmitryduev/kowalski}, an open-source system used internally at Caltech to store the ZTF alert and light curve data together with external catalogs and access those through a standardized API \citep{DuMa2019}.

We used Kowalski to efficiently feed the feature computation pipeline \citep{CoBu2020b} with the ZTF light curve data and store the results. Additionally, the (versioned) classifier predictions have been stored in a dedicated database that fed the active learning process described in Section 4.


\section{Classification scheme} 

Astronomical ground-based light curve data are usually sparse, unevenly sampled, and heteroskedastic, and ZTF is no exception to this general rule. A variable object classification framework must tackle these challenges. First of all, the input image data used to generate the light curves are affected by a broad range of factors such as the weather, the observability of fields, and the cadences of different sub-surveys within ZTF. In addition, the accuracy of the photometry decreases for fainter objects. The result is that objects belonging to the same class will have different noise levels and appear different to the classifier. 

The second problem is that for some types of variable objects, a light curve in a single filter is insufficient for correct classification. Frequently, additional observations are needed in a different bandpass, either optical, infrared, radio or high-energy. In some cases even this is insufficient, and the intrinsic luminosity must be known (e.g. by using the distance from, say, Gaia parallax). Evidently, these data are not always available for all ZTF sources making the external data to be used by the classification framework inhomogeneous and potentially biased towards specific object subsets.

Further, there are several challenges specific to particular classes of variable objects. Some types of objects are more abundant than others so that one has to frequently deal with very imbalanced data sets, with class examples ranging from hundreds of thousands down to just a handful. In addition, the source taxonomy\footnote{taxonomy: a scheme of classification} adds to the challenge as classes can be overlapping. For example, an accreting white dwarf -- red dwarf binary (a cataclysmic variable) can be both outbursting (e.g., a dwarf nova) and eclipsing, or a pulsating star (Cepheid) can be in an eclipsing binary system \citep[e.g.,][]{pietrzynski2010}. This is often caused by the class definitions being a mix of \textit{phenomenological} and ``\textit{ontological}'' (or intrinsic) characteristics of sources.


To tackle these challenges, we employ a hierarchical approach to classification and use a set of \textit{independent binary classifiers}, each of which categorizes the input data set into two groups (e.g., whether or not an object belongs to some class A).

The main advantage of this approach is significantly greater flexibility as compared to the typically used multi-class classifiers, where an object is assumed to have a single correct label of many, or multi-label classifiers, where a single system outputs probabilistic predictions of object class membership for multiple classes at once. 
If the performance on a particular class is deemed insufficient, retraining the classifier with new training data (or employing a different architecture) does not affect the system performance on other classes. Adding new types of variable objects is straightforward and also does not affect other classifiers. As ZTF continuous operations and the temporal baseline and number of epochs increases, new types of variables become detectable, which only requires new classifiers to be added, instead of having to rebuild an entire multi-class or multi-label classifier.

Another advantage is more flexibility for the end-user.
Depending on the (astrophysical) class and the scientific goal, requirements for \textit{completeness}\footnote{As quantified by \textit{recall} or \textit{true positive rate}, i.e. how many relevant items are selected by the classifier.} and \textit{purity}\footnote{As quantified by \textit{precision}, i.e. how many items selected by the classifier are relevant.} can be very different. This trade-off is easier to interpret when using binary classifiers. For example, even though our classifiers are completely independent, they are conceptually organized in a hierarchy so that the ``upstream'' classifiers (high-level classes encompassing a broader range of objects, which are typically trained on larger collections) may be used to increase the sample purity for the ``downstream'' classifiers.

If a classifier is trained on two specific types, the results can be erratic when it is confronted with out-of-distribution objects. The binary classifiers we are using - those that separate their inputs into a given type versus everything else - help alleviate such a problem, making them more robust by allowing the user to impose thresholds along multiple dimensions simultaneously.

Finally, our approach implicitly allows for anomaly detection (for example, the user can select all light curves marked as variable, flaring, and periodic, and not belonging to any other class with high confidence).

These benefits come at a price: the main disadvantage in our approach is that is computationally expensive, both at training and for inference: one would need to train, tune, evaluate, and then use for inference a large number of models instead of a single one.

\begin{figure}
    \centering
    \includegraphics[width=8.5cm]{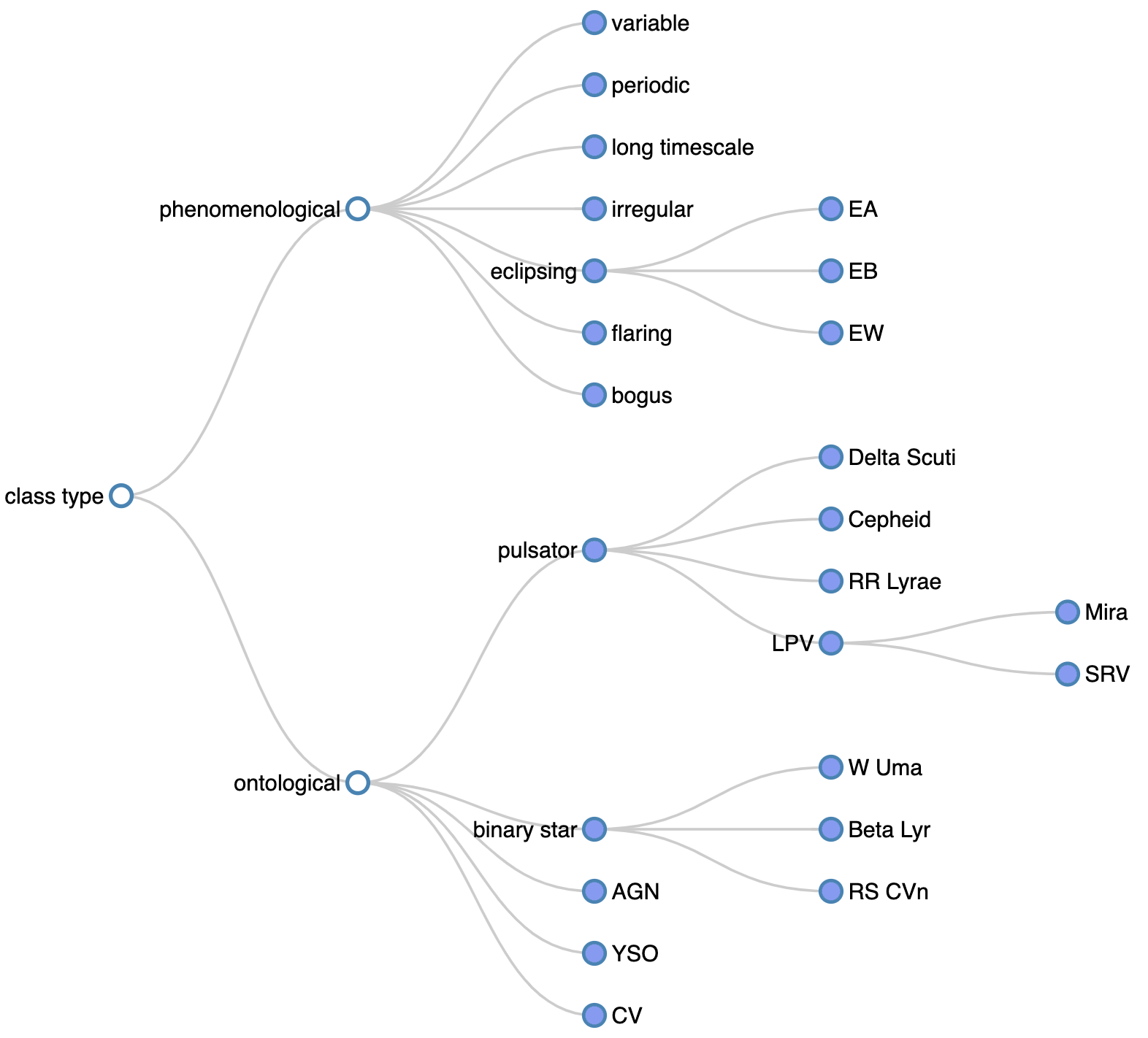}
    \caption{Conceptual hierarchical classification tree of independent binary labels/classifiers used in this work. The filled circles indicate labels for which classifiers were trained.}
    \label{fig:class_tree}
\end{figure}

We organize our labels/classes and the corresponding classifiers into two conceptual groups -- \textit{phenomenological} and \textit{ontological} (see Fig. \ref{fig:class_tree}).\footnote{The figure was generated using the \texttt{tdtax} library,\\ \url{https://github.com/profjsb/timedomain-taxonomy}}

The classifiers of the first group characterize each ZTF object according to the \textit{phenomenological} properties of the corresponding ZTF light curve, e.g. is the object variable, periodic, flaring, eclipsing, etc. The classifiers may act as high-level filters allowing the end-users to efficiently identify objects of interest without imposing a detailed classification scheme. The aim of the ``\textit{phenomenological}'' classifiers is to be as complete and unbiased as possible. Therefore, these classifiers do not use any external data for the classification to avoid biases and enable independent analyses.

Our second group of classifiers - ``\textit{ontological}'' - is geared towards the categorization of specific types of variable objects based on as much information as is available for a particular object. 
The result can then be used to easily obtain a large, pure sample of that particular type of variable. Alternately, by also including lower-scoring examples one can use the result as the input for specialized pipelines to discover new sources (e.g. fitting eclipsing binary light curves with binary star models).
We note that these classifiers use features from external catalogs in addition to ZTF data and are therefore prone to non-ZTF-specific biases.

\subsection{Machine learning algorithms}
To automatically classify all ZTF light curves, we use supervised machine learning algorithms. Supervised machine learning algorithms  ``learn'' mappings between the input and the output spaces from a training set (for which both the input and output are known). This is achieved by solving an optimization problem of minimizing a loss function that quantifies the gap between prediction and ground truth. How this mapping is constructed depends on the machine learning algorithm and can be tuned by changing the values of ``hyperparameters''.
In this work, we use two different types of supervised machine learning methods.

In the first case, we employ deep learning methods, referred to hereafter as the deep neural networks (DNN). DNN are universal function approximators that can learn arbitrary mappings between the input and the output spaces. The network's output is produced using multiple simple non-linear transformations organized in interconnected ``layers''. The networks are typically ``trained'' by alternating forward and backward passes -- computing a prediction and then updating the trainable transformation parameters (weights and biases) to decrease the loss function. Neural networks are extremely flexible, with the number of layers and the number of nodes per layer as some of the most important hyperparameters.

The second type of classifiers are gradient boosted decision tree classifiers \citep{friedman2001}, implemented in {\sc XGBoost} \citep{chen2016}. This type of classifier is based on a series of decision trees used as weak learners. They have real-valued outputs that can be added together and used to implement splits. The trees are gradually grown, with the additions being weighted such that the classifier performance improves on the earlier values. The growth is carried out in a greedy fashion, based on purity scores and minimization of the loss function. The thresholds for the accumulating values, the number of trees, etc., can be used as hyperparameters making this method extremely adaptable and general. As in random forests \citep{Ho1995}, random subsets of features and the data are used per iterations.

\section{Data set assembly and classifier training}\label{sec:active_learning}
As we noted above, labeled light curves from a multitude of previous and current surveys are available. However, we decided not to blindly use those because that would inevitably introduce survey-specific biases. Instead, we employed an active-learning approach of alternating between data labeling and classifier training with subsequent sampling of their predictions, both confident ones and those near the decision boundary.


\begin{figure*}
    \centering
    \includegraphics[width=\textwidth]{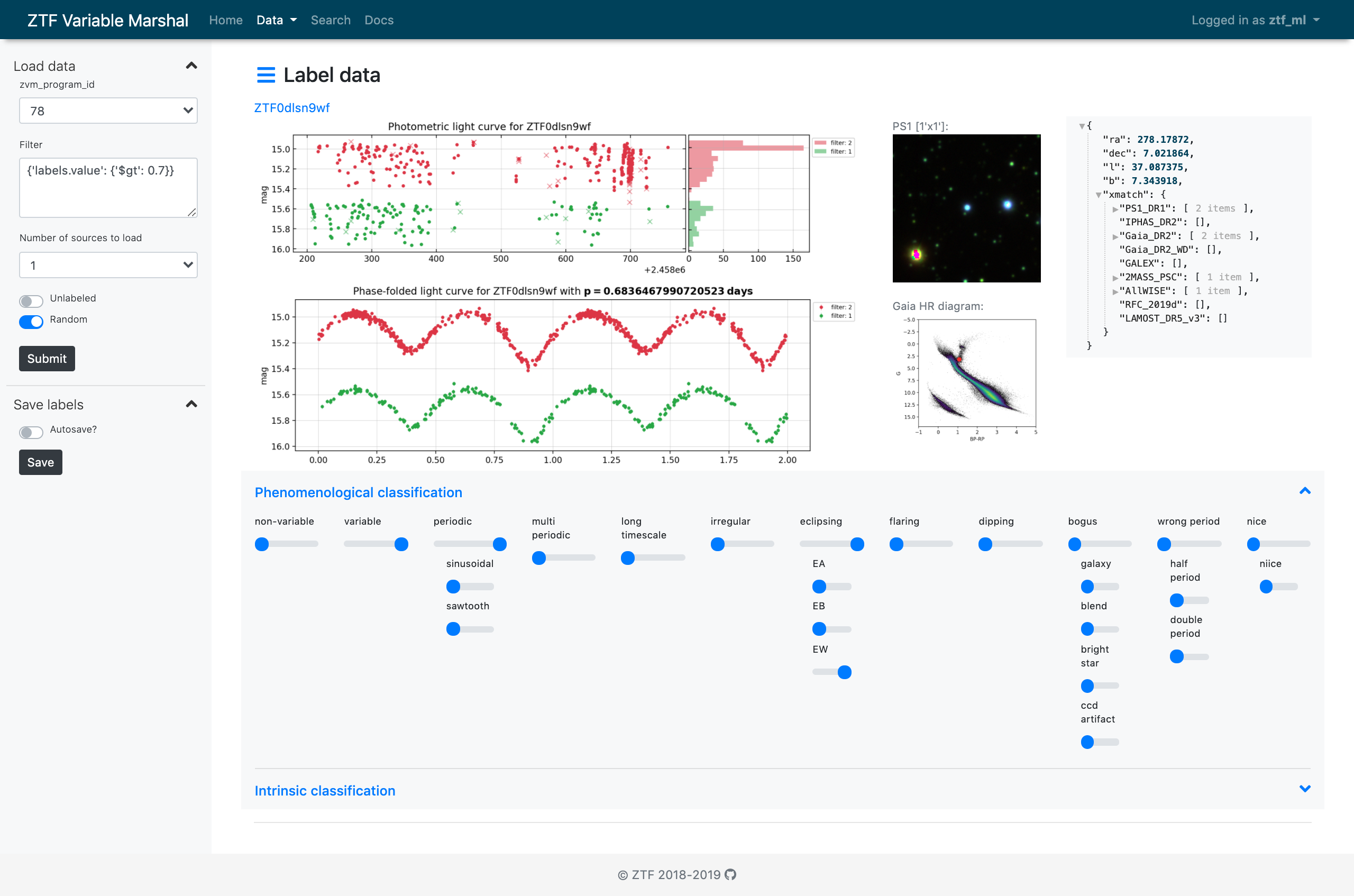}
    \caption{Labeling interface of the ZTF Variable Marshal.}
    \label{fig:zvm}
\end{figure*}

To streamline data labeling, we have built a dedicated extension of the ZTF Variable Marshal\footnote{\url{https://github.com/dmitryduev/ztf-variable-marshal}}, an open-source web application 
for interactive exploration, analysis, and annotation of the ZTF variable sources (see Fig. \ref{fig:zvm}). The API-driven interface displays the ZTF light curves for each filter per object, along with an additional set of light curves that are phase-folded to any period (or periods) associated with an object. As additional information, the location on the Gaia observational HR diagram and a Pan-STARRS image cutout are displayed. Labels can be assigned using a set of range sliders representing the class labels. The slider values are quantized to 0, 0.25, 0.5, 0.75 and 1, to enable a human scanner to indicate how certain they are of their classification, the information that can be used in classifier training. Being part of the ZTF Variable Marshal, all interactions with the interface can be carried out programmatically via API calls.

Contrary to a common misconception, data labeling is actually a job for highly-skilled, trained professionals that takes most of the time and is one of the most important parts of the work to build any successful ML system. We started with multiple experts performing classification using different user accounts, 
but later moved to regular multi-expert classification sprints that used a single user account. This approach proved to be superior as it effectively averaged input from multiple experts and minimized the number of mistakes while labeling\footnote{This is somewhat similar to the agile software development technique of pair programming}.

\begin{figure}
    \centering
    \includegraphics[width=\columnwidth]{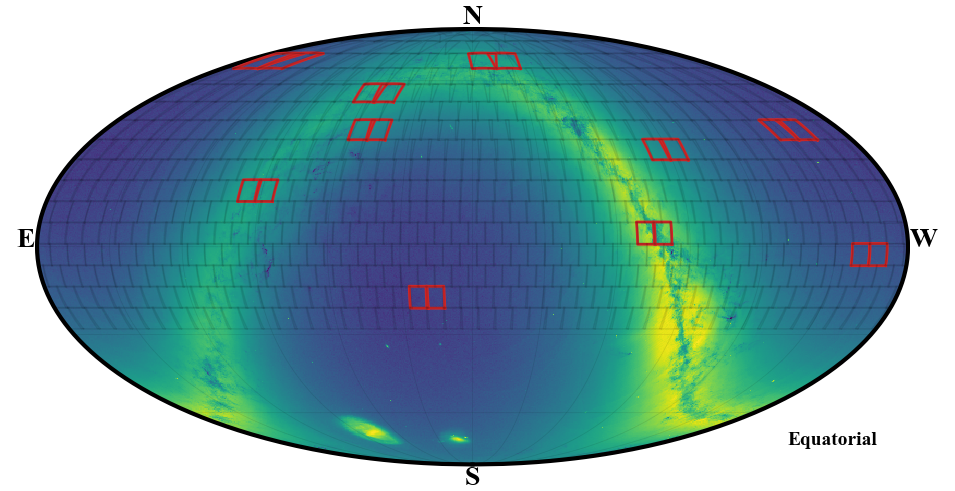}
    \caption{The location on sky of the 10 pairs of ZTF-fields (red) we used for testing our pipeline. The figure uses Equatorial coordinates and a Mollweide projection. The background shows the stellar density according to Gaia DR2 using a logarithmic scaling.}
    \label{fig:testfields}
\end{figure}

To test our pipeline during development, we selected a subset of the ZTF data. In order to obtain a representative set, we chose ten pairs of ZTF-fields, taking into account the RA and Dec, and the Galactic latitude, see Fig. \ref{fig:testfields} and Table \ref{tab:testfields}. These fields contain a diverse range of Galactic environments and also span a range of different cadences and total number of epochs. We use the $g$, $r$, and $i$ band light curve of these fields, a total of $\approx$34 million.

In order to explore the data and build an initial training sample, we visually inspected two sets of random and non-variable light curves (from a field in the test-set) by simply selecting light curves identified as outliers in the IQR-magnitude phase space. We visually inspected $\approx$2000 candidates with excess IQR values and $\approx$1000 random light curves without any excess in the IQR score. This showed that there were many ``bogus'' variable candidates (blended stars, diffraction spikes) present in the sample of excess IQR values, while there were also variable objects (mainly eclipsing binaries) in the sample without any IQR excess. After further experimentation with additional features, their combinations, and unsupervised ML clustering algorithms, we concluded that \textit{no simple selection method could be found, which yields a sample that is sufficiently clean and simultaneously representative of the data}.

Instead, to efficiently increase the size of the training sample while keeping it representative, we used the input from human scanners to build a 'seed' variable/non-variable (``vnv'') classifier. We used the labels obtained from the initial scanning effort to build a simple classifier (see below) and inspected random samples of $\approx1000$s low, medium and high scoring variable candidates, which we labeled and added to the training sample. 

In addition, we added a large diverse set of visually classified examples, which were under investigation by various ZTF team members, including cataclysmic variables (outbursting and non-outbursting from SDSS \citep{szkody2011}, CRTS \citep{drake2014,breedt2014}, PTF (Groot priv. com.) and ZTF \citep{szkody2020}, and, RR Lyrae, eclipsing binaries, Delta Scuti, Cepheids, Long Period Variables, variable YSOs and AGN. 

For each class (see Fig. \ref{fig:class_tree}), we trained a dedicated classifier that was executed on the full (unlabeled) data sample from the 20 test fields ($\approx34$ million light curves). Similarly to the seed ``vnv'' classifier, the predictions were randomly sampled for low and high-scoring candidates as well as ``abstained'' examples (meaning that their scores were close to 0.5 -- the classifier decision threshold used at training), and the resulting sets were inspected and labeled by human experts. This process was repeated multiple times over.

Next, we applied this set of classifiers to the stars from the CRTS sample of periodic variables \citep{drake2014}. We visually inspected all objects for which the prediction did not match the CRTS label. We then added the CRTS labels to the training sample, which we used to train the next set of classifiers.

\begin{figure}
    \centering
   \includegraphics[width=8cm]{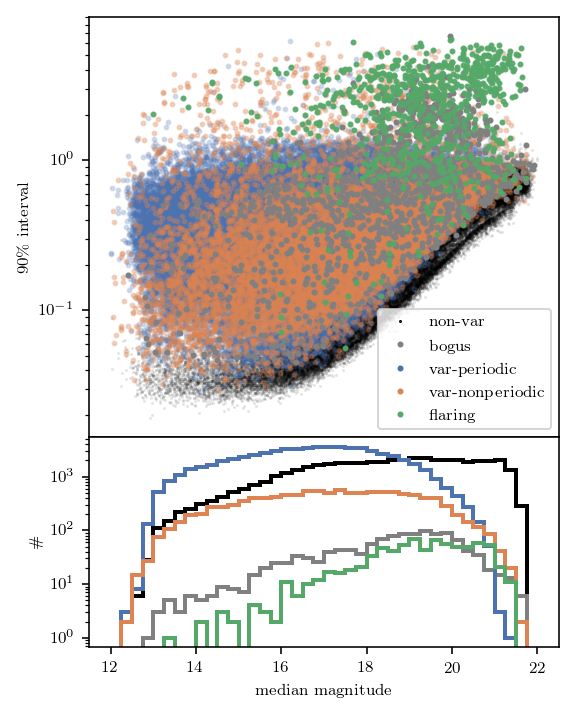}
    \caption{The parameter distribution of the training set with high-level classes indicated with different colors. The top panel show the 90\% interval (a measure of amplitude) and the median magnitude. Objects with a large amplitude are often variables, but so are many ``bogus'' light curves (often artifacts due to bright stars). This figure also show that there is no clear separation between variables and non-variables. The bottom panel show the median magnitude distribution of high-level classes.}
    \label{fig:training_set}
\end{figure}

\begin{figure*}
    \centering
    \includegraphics[width=\textwidth]{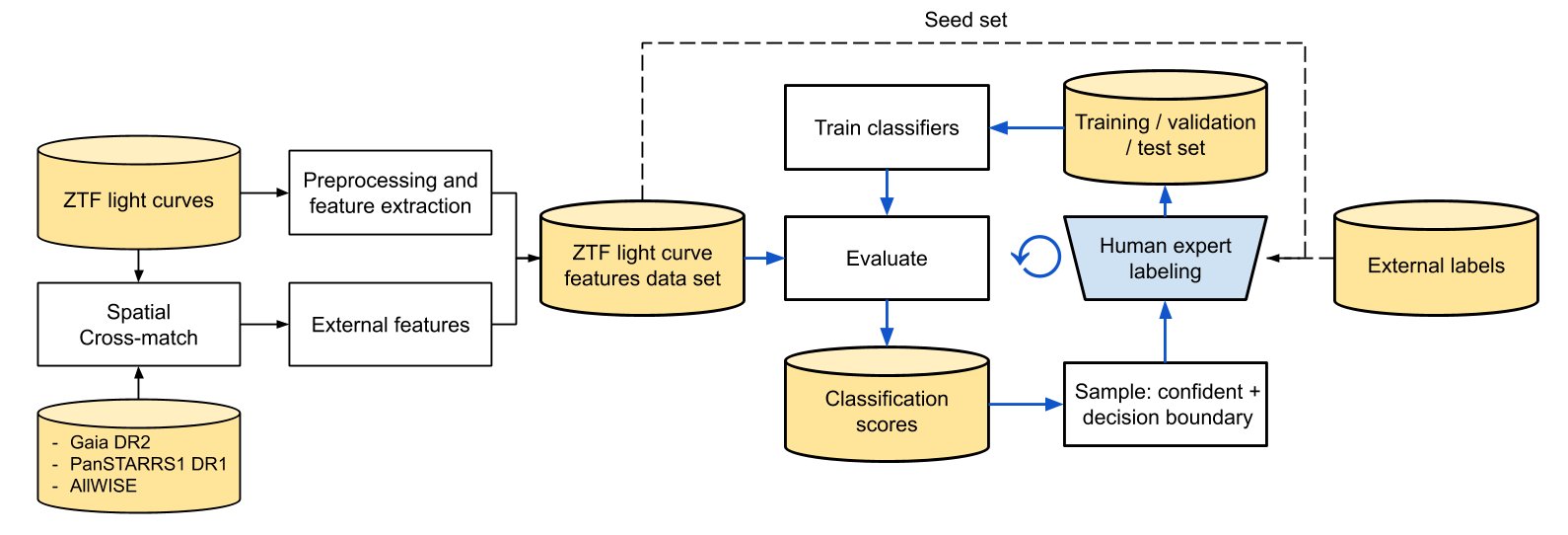}
    \caption{Flowchart of the workflow. Features are extracted from the pre-processed ZTF light curves and combined with external features from Gaia DR2, PanSTARRS1 DR1, and AllWISE via a spatial cross-match. The resulting feature data set is sampled for a small ``seed'' set for human expert labeling. Externally-labeled data are inspected by the experts as well. The blue arrows show the active learning process for iteratively building the training set and improving the classifier performance. The labeled examples are assembled into a training/validation/test set that is used for classifier training. The phenomenological classifiers use only the ZTF-based features in the process, while the ontological ones additionally use the external features. The resulting set of trained classifiers is evaluated on the full light curve features data set. The resulting (versioned) scores are stored in a database and sampled both for confident and near-the-decision-boundary predictions and passed for labeling to begin a new active learning cycle.}
    \label{fig:workflow}
\end{figure*}

Finally, we performed several more rounds of the train-infer-sample-label active learning process. As expected, with each completed cycle, we observed a gradual improvement of the classifier performance (as determined from a ``hold-out'' set).
We stress that the resulting labeled data set is very much \textit{a living entity}. 

Several characteristics of the training set as of the time of writing (internal tag \texttt{d11}) are shown in Fig. \ref{fig:training_set} \edit1(with the total number of objects per class in Table \ref{tab:test_set_performance}). This shows that there are approximately the same amount of variables and non-variables. Note that magnitude distribution is different for the high-level classes. This is partially due to the intrinsic distribution of objects, but mostly due to selection biases in the training set. 

From the training data, we separated a few sets of $\approx$100 objects each for the ontological classes, the 'gold' samples. The light curves in these sets were selected as very easy to classify examples of those particular classes. These sets are meant as verification sets, to be used as a `sanity check' for both the phenomenological and ontological classifiers.

The workflow described above is summarized in Fig. \ref{fig:workflow}.

\subsection{Training process}
\subsubsection{DNN}

\begin{figure}
    \centering
    \includegraphics[width=8cm]{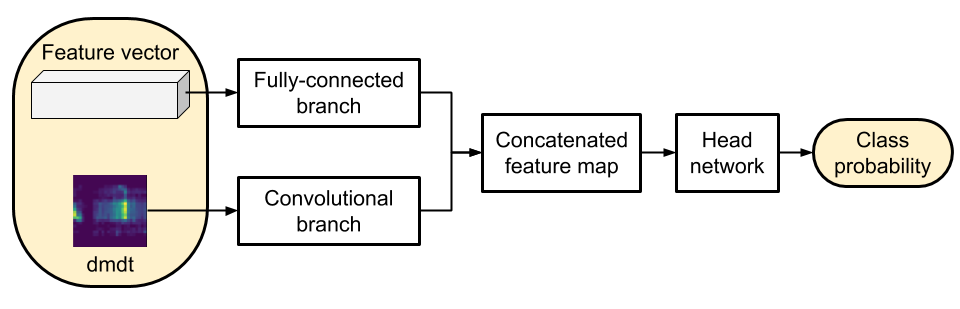}
    \caption{Schematic of the conceptual DNN architecture.}
    \label{fig:dnn-concept}
\end{figure}

\begin{figure}
    \centering
        \subfigure[Architecture used in production]{\includegraphics[height=13cm]{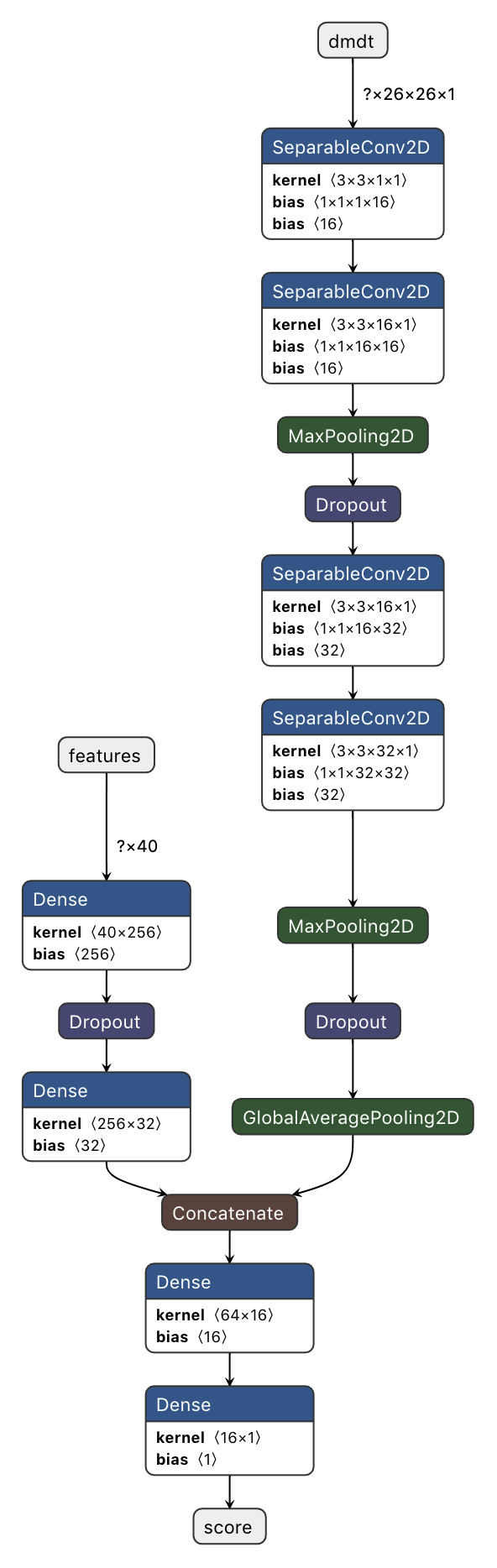}}\quad
        \subfigure[Example of a more complicated architecture]{\includegraphics[height=13cm]{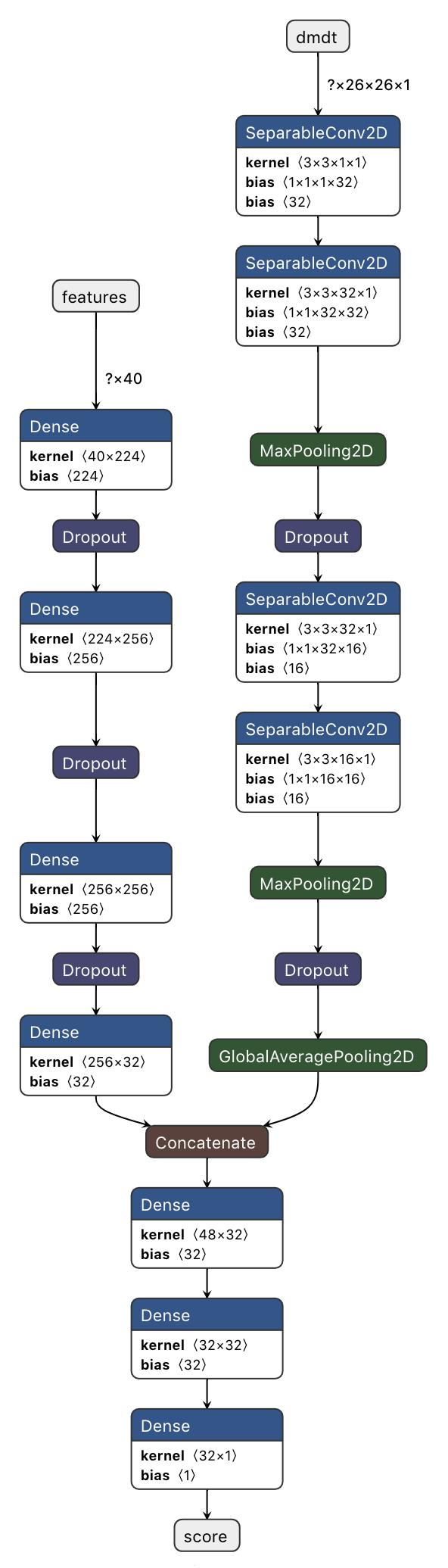}}\quad
    \caption{Best-performing DNN architectures. Panel (a) shows the architecture used in production phenomenological classifiers. The same architecture is used for the ontological classifiers with the difference being the input feature vector size (69 vs 40). Panel (b) shows an example of a more complicated architecture that tends to show higher variance compared to (a).}
    \label{fig:dnn-architecture}
\end{figure}

When building the DNN classifiers, we had to explore a vast hyperparameter space. Figure \ref{fig:dnn-concept} illustrates the conceptual DNN architecture that we iterated on:
\begin{itemize}
    \item The phenomenological classifiers use the pre-computed light curve features and \textit{dmdt} histograms as input. The ontological classifiers additionally use the external features.
    \item A dense neural network containing multiple fully-connected layers is used to process the features.
    \item A convolutional neural network is used to process the \textit{dmdt}'s.
    \item The resulting feature maps are fused and passed through a fully-connected ``head'' network that outputs the final classification score.
\end{itemize}

The classifiers were implemented using \texttt{TensorFlow} software and its high-level \texttt{Keras} API \citep{tensorflow2015-whitepaper, chollet2015keras}. We used the binary cross-entropy loss function, the Adam optimizer \citep{2014arXiv1412.6980K}, a batch size of 64, and a $81\%/9\%/10\%$ training/validation/test data split with a fixed random seed for reproducibility. 
The input features were normalized; the same norms were used for all classifiers. We did class balancing of the training sets for the classifiers with a small number of positive examples and used all available data for the classifiers with a large number of available examples. In the first case, the classifier performance was checked on the originally dropped negative examples and the small number of misclassifications (typically on the order of $1-3\%$) were added to the training set.
The training data were weighted per class. The class weights were further adjusted to balance precision (purity) and recall (completeness).
We used the standard techniques to achieve the best performance such as learning rate reduction on a plateau and early stopping based on validation loss.

For the initial ``seed'' \textit{vnv} classifier, we used a simple architecture that followed the schematic in Fig. \ref{fig:dnn-concept} and demonstrated satisfying performance, with a minimal number (chosen arbitrarily) of fully-connected and convolutional layers. As we expanded the data sets and added more classifiers, we ran several rounds of hyperparameter tuning using the \texttt{keras-tuner}\footnote{\url{https://github.com/keras-team/keras-tuner}} library \citep{omalley2019kerastuner}. The following hyperparameters were tuned:

\begin{itemize}
    \item Inclusion of the fully-connected branch in the architecture or not (provided the convolutional branch is included)?
    \begin{itemize}
        \item Number of layers (from 1 to 4) and neurons therein (from 32 to 512 with a step of 32)
    \end{itemize}
    \item  Inclusion of the convolutional branch in the architecture or not (provided the fully-connected branch is included)?
    \begin{itemize}
        \item Number of filters (from 16 to 64 with a step of 16), their size (3x3, 5x5, or 7x7) and type (regular or separable convolution)
        \item Flattening the output of the last convolutional block or use global average pooling instead
    \end{itemize}
    \item Number of layers and neurons in the head network (from 0 to 3)
    \item Dropout rates (from 0.15 to 0.55 with a step of 0.1)
    \item Activation functions (ReLU, leaky ReLU, sigmoid, tanh)
    \item Initial learning rate (from 1e-4 to 1e-3 with a step of 1e-4)
\end{itemize}

Several best-performing architectures were evaluated on the test sets described below in Sec. \ref{sec:classifier_performance}. As expected, the more complicated architectures tended to show higher variance\footnote{In ML, variance is usually defined as the error from sensitivity to small fluctuations in the training set.} so for production, we selected the simplest architecture that yielded the most robust performance in most cases (see Fig. \ref{fig:dnn-architecture}). The architecture includes both the fully-connected and the convolutional branches confirming that using \textit{dmdt}'s indeed improves classifier performance. It uses separable convolutions \citep[see e.g.][]{2016arXiv161002357C}, ReLU activation functions for all hidden trainable layers and a sigmoid activation function for the output layer that produces a score from 0.0 to 1.0. Dropout layers with a rate of 0.25 are used for regularization.

\subsubsection{XGBoost}

Similar to DNN, XGBoost has a large number of hyperparameters. These can be categorized as general parameters, tree boosting parameters, learning parameters, etc. A thorough hyperparameter tuning is generally not possible, and, indeed, not practical. Various methods are adopted to find near-optimal values for some of the parameters that should be tuned. Some of the critical parameters are:
\begin{itemize}
\item $max\_depth$: this indicates the depth of the tree, with greater depth indicating more complex models, in turn implying models that are more prone to overfitting,
\item $min\_child\_weight$: this is a parameter that determines when further partitioning of a tree will stop. Larger numbers indicate a more conservative approach,
\item $subsample$: this determines the fraction of the data that the boosting algorithm will use at each boosting iteration,
\item $colsample\_bytree$: this is a counterpart to $subsample$ but pertaining to the columns. In other words, it is the number of features that will get used in each tree,
\item $eta$: this is the learning rate and is applied after every boosting step. 
\end{itemize}

All of these parameters affect tree boosting. We tuned these parameters, and, since we use all the available data which is very unbalanced, we tuned one more viz.
\begin{itemize}
\item $scale\_pos\_weight$: this parameter decides if one of the classes needs to be given extra weight while fitting because it has fewer samples. Given the way XGBoost determines the splits using its complex parameters, even for unbalanced classes, one is often fine with leaving this parameter set to one.
\end{itemize}

We started with $scale\_pos\_weight$, giving it four choices viz. [1, CR/2, CR, 2*CR] where CR is the ratio of samples belonging to the two classes.
Then we tuned $max\_depth$ (from 3 to 7 at a spacing of 2) and $min\_child\_weight$ (from 1 to 5 at a spacing of 2) simultaneously, sampling the grid at nine points. Then we sampled near the optimal point at a spacing of 1, thus covering [2,8] for $max\_depth$ and [1,6] for $min\_child\_weight$. This was followed by similar simultaneous tuning of $subsample$ (from 0.6 to 1.0 at the spacing of 0.2) and $colsample\_bytree$ (from 0.6 to 1.0 at a spacing of 0.2). Here too we did a second round of tuning near the optimal point at a spacing of 0.1, resulting in a cover of [0.5,1.0] each for $subsample$ and $colsample\_bytree$. This was then followed by tuning eta at the values [0.3, 0.2, 0.1, 0.05]. Then we went back to $scale\_pos\_weight$ to ensure that the value we had determined at the start was still the best value. In all cases, $scale\_pos\_weight$ was 1 or close to one.

We did two sets of classifications with different inputs sets of features (1) for the phenomenological classes we used 40 features determined from the ZTF light curves alone (see Table \ref{tab:ZTFfeatures}), and 
(2) for the ontological classes, we used the 29 external features from AllWISE, Gaia, and Pan-STARRS along with the 40 ZTF features as with the DNN classifiers (see Table \ref{tab:Externalfeatures}). Metrics from these runs are given in Table \ref{tab:test_set_performance}.

\subsection{T-SNE analysis of the training set}

\begin{figure}
    \centering
   \includegraphics[width=8cm]{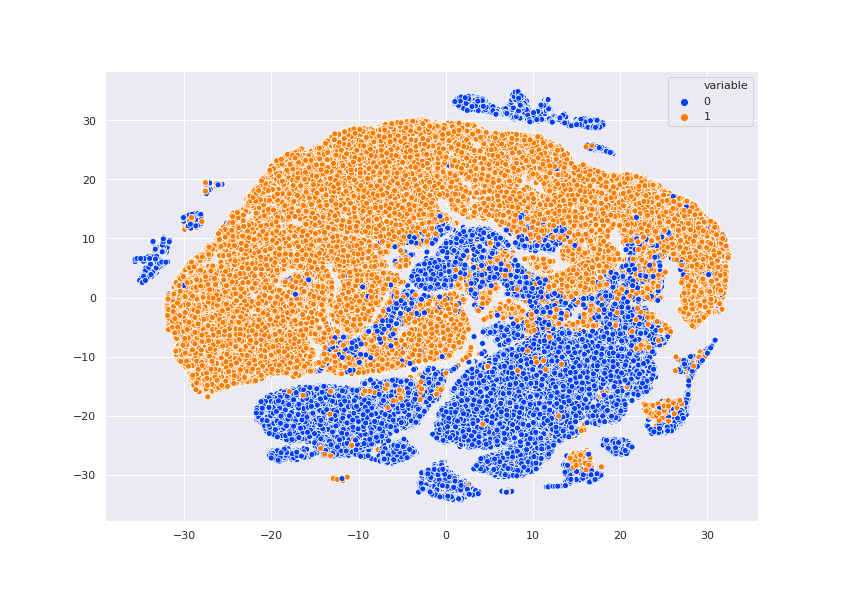}
   \includegraphics[width=8cm]{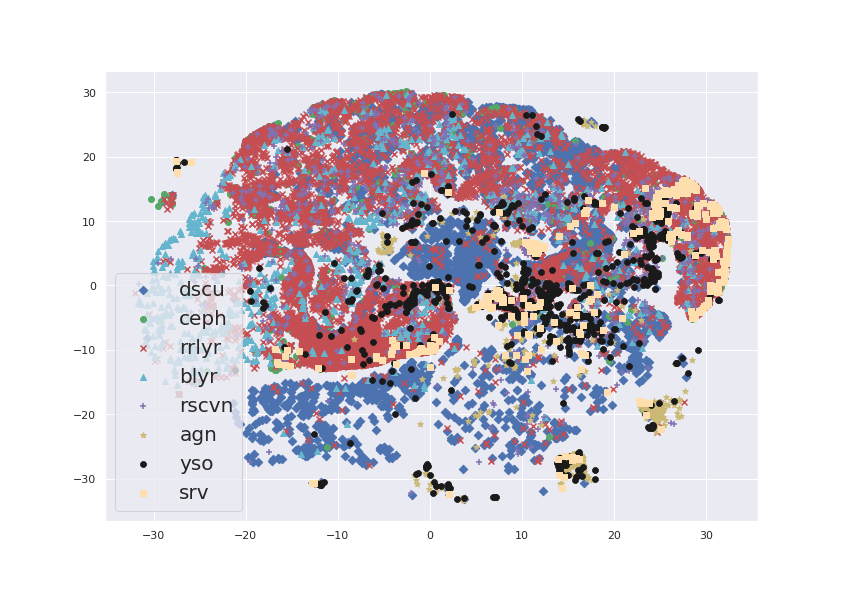}
    \caption{An overview of the training set using t-SNE. Top: variables and non-variables, Bottom: Leaf-level ontological classes within the set of variables.
    }
    \label{fig:training_set_t_sne}
\end{figure}

As described in Section \ref{sec:active_learning}, we built our training set through a series of iterative steps. Both DNN and XGBoost use a set of features for classifications. We passed these sets of features for our training sample to t-distributed Stochastic Neighbor Embedding \citep[t-SNE][]{vanDerMaaten2008}, a dimensionality reduction technique. t-SNE maps points near each other in a high-dimensional space to its low dimensional counterpart by minimizing KL divergence \citep{kullback1951} between the two probability distributions using gradient descent. 
In Fig. \ref{fig:training_set_t_sne}, we plot variables and non-variables separately and then plot the leaf-level ontological classes by leaving out the non-variables. Many classes are seen to be clustered, but there is also overlap between some others. This is to be expected especially for classes with relatively fewer examples and the overlaps can be used to predict classes with possible ambiguities when running inference on light curves of unknown objects.

\section{Classifier performance}
\label{sec:classifier_performance}

We have tested our classifiers on different labeled sets. The test performance is based on a random split of the training set (10\% of the examples for a given class). The performance on the test set indicates the ability of the classifiers to learn the decision boundaries. Table \ref{tab:test_set_performance} summarizes different metrics of our classifiers on these sets. As can be seen in the table, the performance of the DNN and XGBoost method is similar in most cases, with similar values across the board. However, for XGBoost the performance deteriorates when the class imbalance between two classes is more than a factor of 30, for example for the Beta Lyrae (blyr) class.

\begin{table*}
\begin{tabular}{|l|l|l|l|l|l|l|l|l|l|}
\hline
Class & \# & \multicolumn{2}{c|}{Accuracy} & \multicolumn{2}{c|}{Precision} & \multicolumn{2}{c|}{Recall} & \multicolumn{2}{c|}{F1 Score} \\ \hline
        &               &       DNN     &       XGB     &       DNN     &       XGB     &       DNN     &       XGB     &       DNN     &       XGB\\ \hline
e       &       44721   &       0.94    &       0.95    &       0.9     &       0.92    &       0.93    &       0.95    &       0.92    &       0.93\\
ea      &       819     &       0.94    &       1       &       0.91    &       1       &       0.87    &       0.02    &       0.89    &       0.03\\
eb      &       950     &       0.88    &       0.99    &       0.86    &            &       0.74    &       0       &       0.8     &       \\
ew      &       39079   &       0.94    &       0.95    &       0.91    &       0.92    &       0.89    &       0.93    &       0.9     &       0.92\\
fla     &       829     &       0.97    &       1       &       1       &       0.84    &       0.87    &       0.82    &       0.93    &       0.83\\
i       &       1842    &       0.93    &       0.99    &       0.92    &       0.79    &       0.84    &       0.28    &       0.88    &       0.42\\
longt   &       968     &       0.95    &       1       &       0.93    &       0.87    &       0.93    &       0.38    &       0.93    &       0.53\\
pnp     &       64910   &       0.95    &       0.95    &       0.95    &       0.95    &       0.96    &       0.96    &       0.96    &       0.95\\
vnv     &       78083   &       0.97    &       0.98    &       0.99    &       0.98    &       0.97    &       0.98    &       0.98    &       0.98\\
\hline
agn     &       608     &       0.98    &       1       &       0.94    &       0.94    &       0.98    &       0.71    &       0.96    &       0.81\\
bis     &       44532   &       0.95    &       0.96    &       0.92    &       0.93    &       0.93    &       0.96    &       0.93    &       0.94\\
blyr    &       836     &       0.89    &       0.99    &       0.8     &       0.46    &       0.81    &       0.9     &       0.81    &       0.61\\
ceph    &       1075    &       0.93    &       1       &       0.88    &       0.76    &       0.89    &       0.92    &       0.89    &       0.83\\
dscu    &       6118    &       0.96    &       1       &       0.92    &       0.96    &       0.93    &       0.97    &       0.93    &       0.96\\
puls    &       18664   &       0.96    &       0.99    &       0.94    &       0.94    &       0.93    &       0.98    &       0.94    &       0.96\\
lpv     &       968     &       0.99    &       1       &       0.97    &       0.88    &       0.99    &       0.79    &       0.98    &       0.84\\
rrlyr   &       10866   &       0.95    &       0.99    &       0.93    &       0.95    &       0.89    &       0.95    &       0.91    &       0.95\\
rscvn   &       1210    &       0.85    &       1       &       0.83    &       0.77    &       0.68    &       0.82    &       0.75    &       0.8\\
srv     &       420     &       0.95    &       1       &       0.88    &       0.81    &       0.98    &       0.69    &       0.93    &       0.74\\
yso     &       849     &       0.99    &       1       &       0.99    &       0.92    &       0.99    &       0.99    &       0.99    &       0.95\\ 
\hline
\end{tabular}
\caption{Test set performance of our classifiers using a score threshold of 0.5. Labeled data set version \textit{d11}. Total number of light curves in the set 124,037. See the appendix for the definition of each of the classes. The first half of the table shows phenomenological classes, the second half the ontological classes. The second column shows the total number of labeled examples of the corresponding class in the set; the classifiers were evaluated on 10\% of those. For the phenomenological classes only features from ZTF data were used (excluding \textit{dmdt} for XGBoost).}
\label{tab:test_set_performance}
\end{table*}

\begin{figure*}
  \centering
  \subfigure[RR Lyrae ab]{\includegraphics[width=0.48\textwidth]{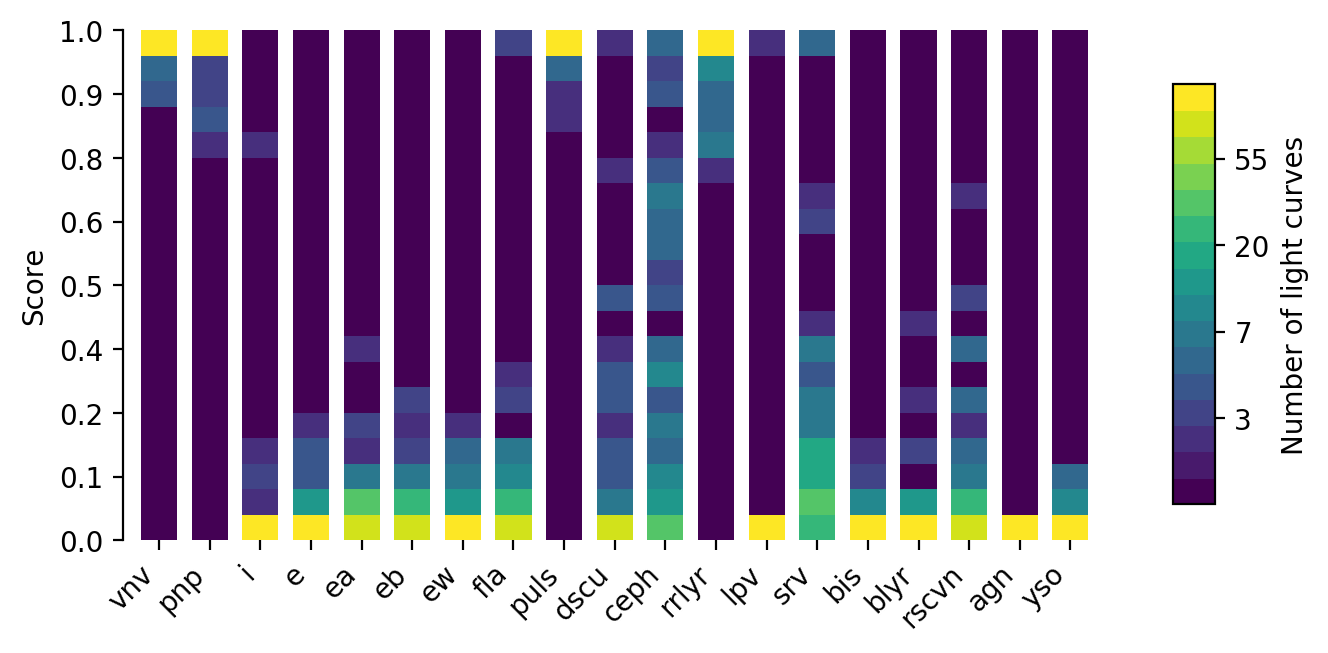}}\quad
  \subfigure[RR Lyrae c]{\includegraphics[width=0.48\textwidth]{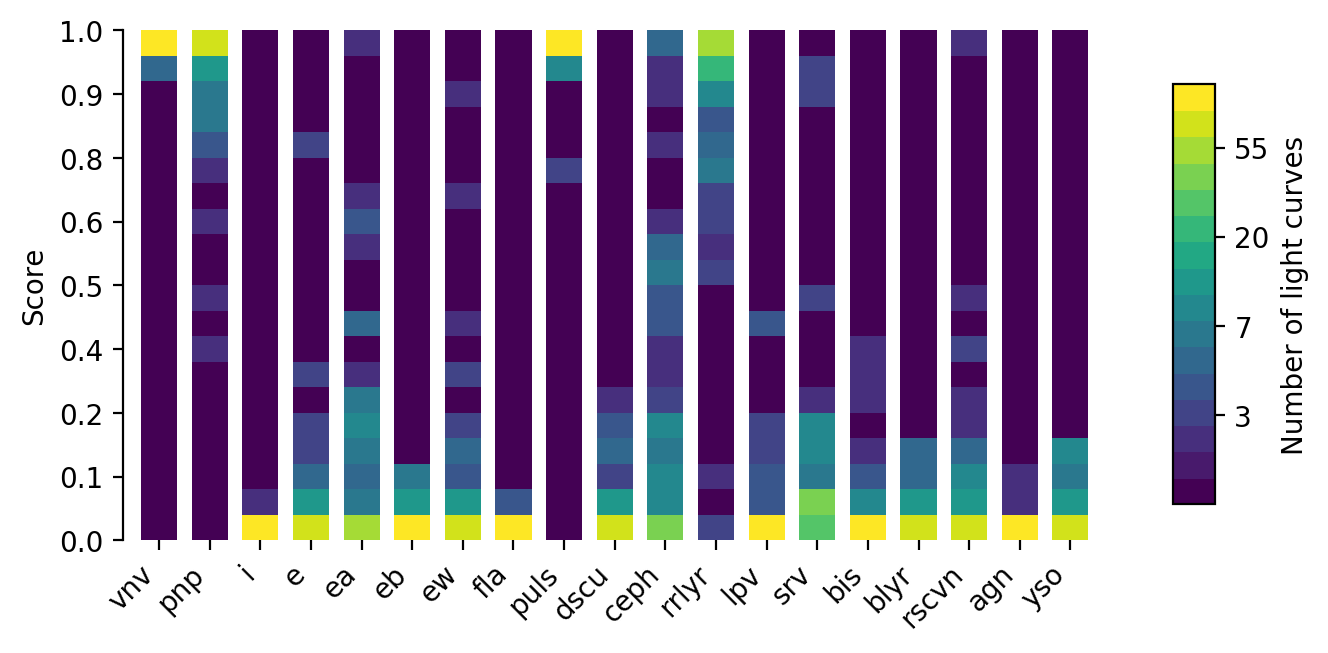}}\quad
  \subfigure[Flaring stars]{\includegraphics[width=0.48\textwidth]{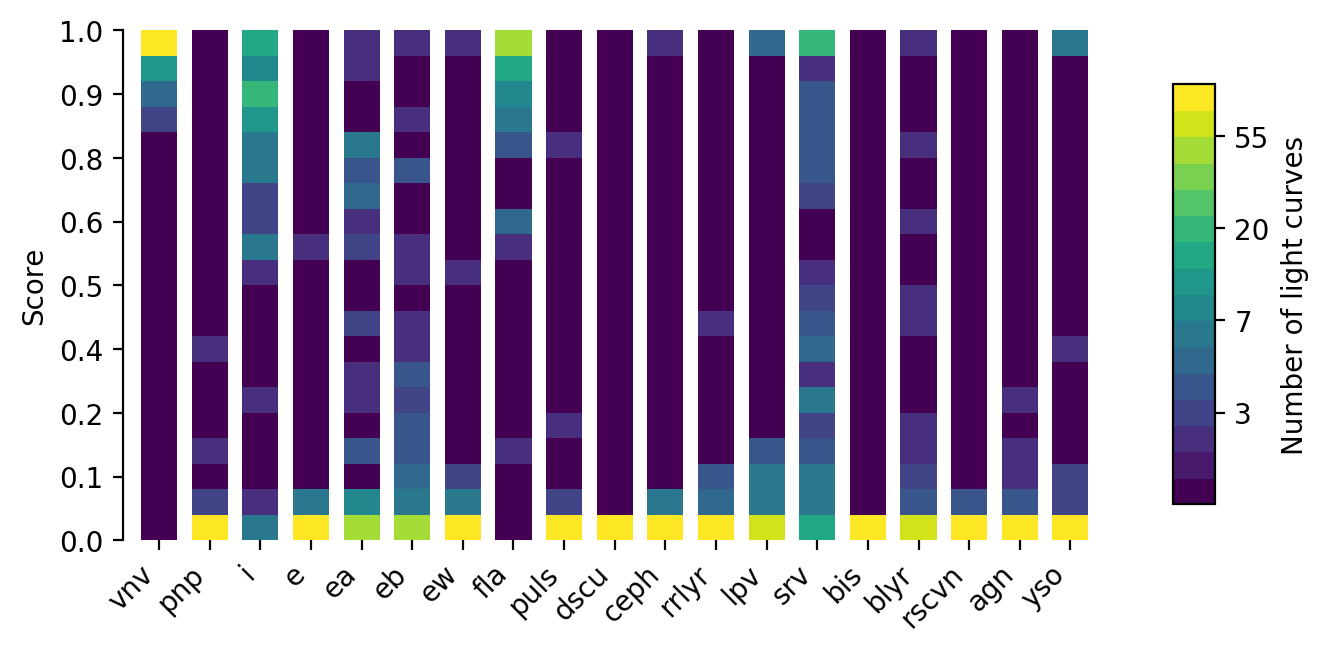}}\quad
  \subfigure[EA]{\includegraphics[width=0.48\textwidth]{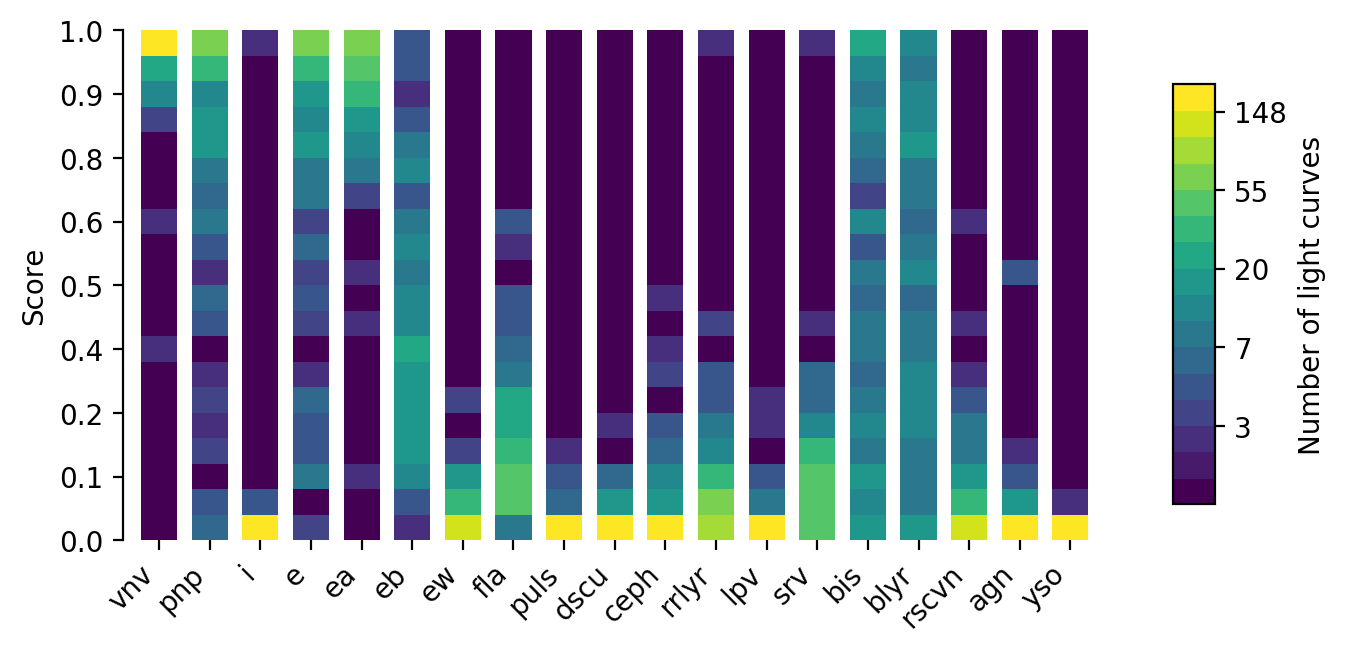}}\quad
  \subfigure[EB]{\includegraphics[width=0.48\textwidth]{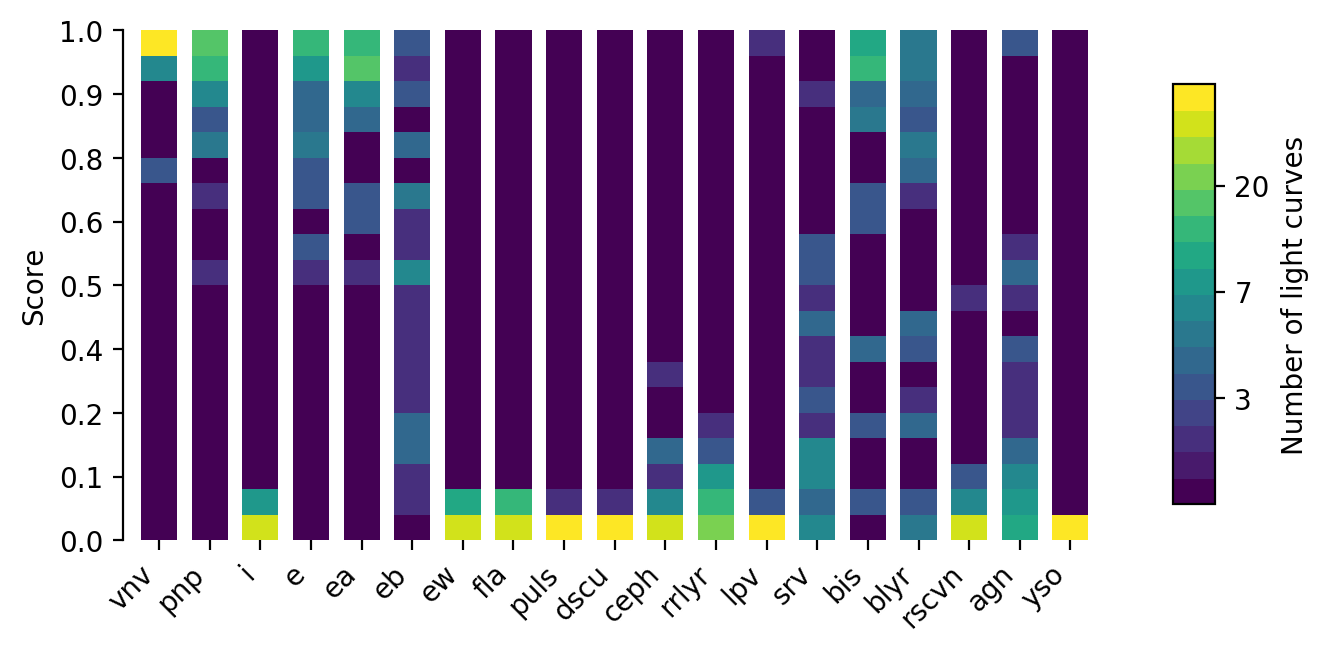}}\quad
  \subfigure[EW]{\includegraphics[width=0.48\textwidth]{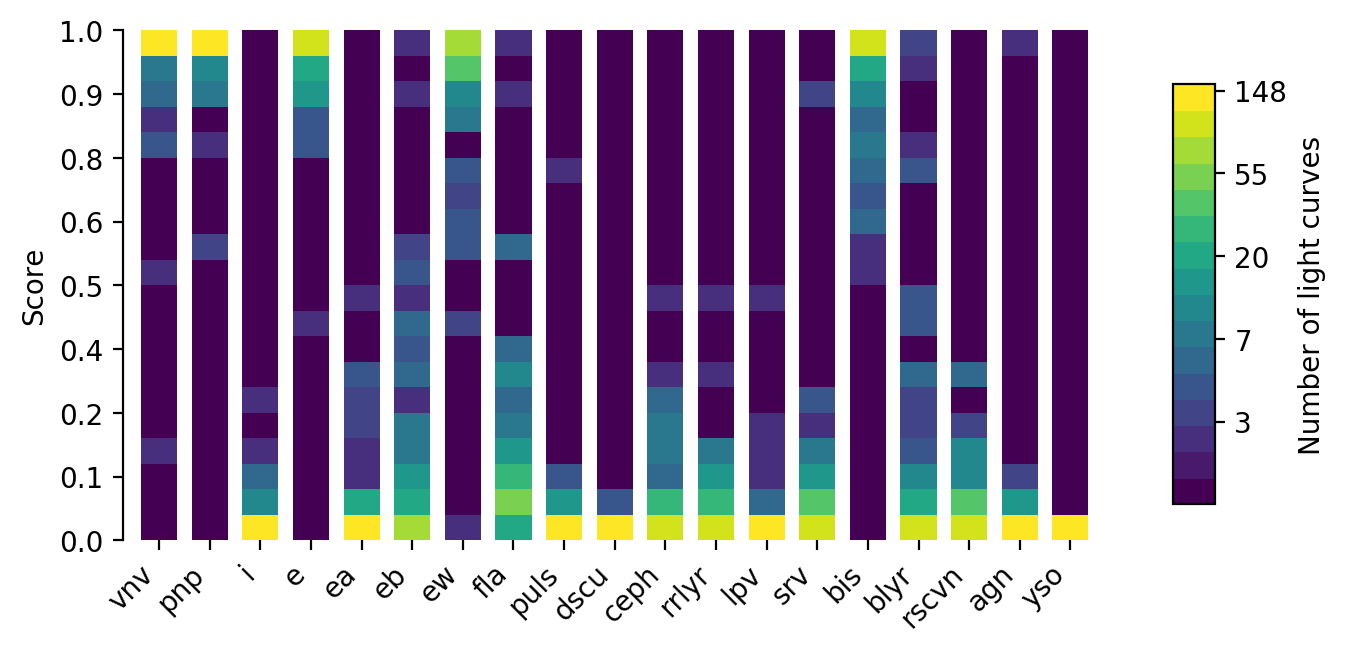}}\quad
\caption{Score distributions color-coded in logarithmic scale of all the DNN classifiers on different object types from the gold set.}
\label{fig:dnn_on_gold}
\end{figure*}

The ``gold'' set performance can be considered a sanity check. The gold sets have been identified by human scanners and are not part of the training sets. These sets contain easy-to-classify objects, and our classifiers should demonstrate excellent performance on them, which is indeed the case (see Fig. \ref{fig:dnn_on_gold}). As can be seen in the figure, most classifiers confidently classify the gold set correctly. There are a few exceptions, for example, the semi-regular variable classifier does not seem to perform as well as the other classifiers (mostly due to a lack of examples). There is also some confusion between some of the binary classes; the difference between EA and EB light curves is subtle, so this confusion is expected. 

\begin{figure*}
  \centering
  \subfigure[RR Lyrae]{\includegraphics[width=0.48\textwidth]{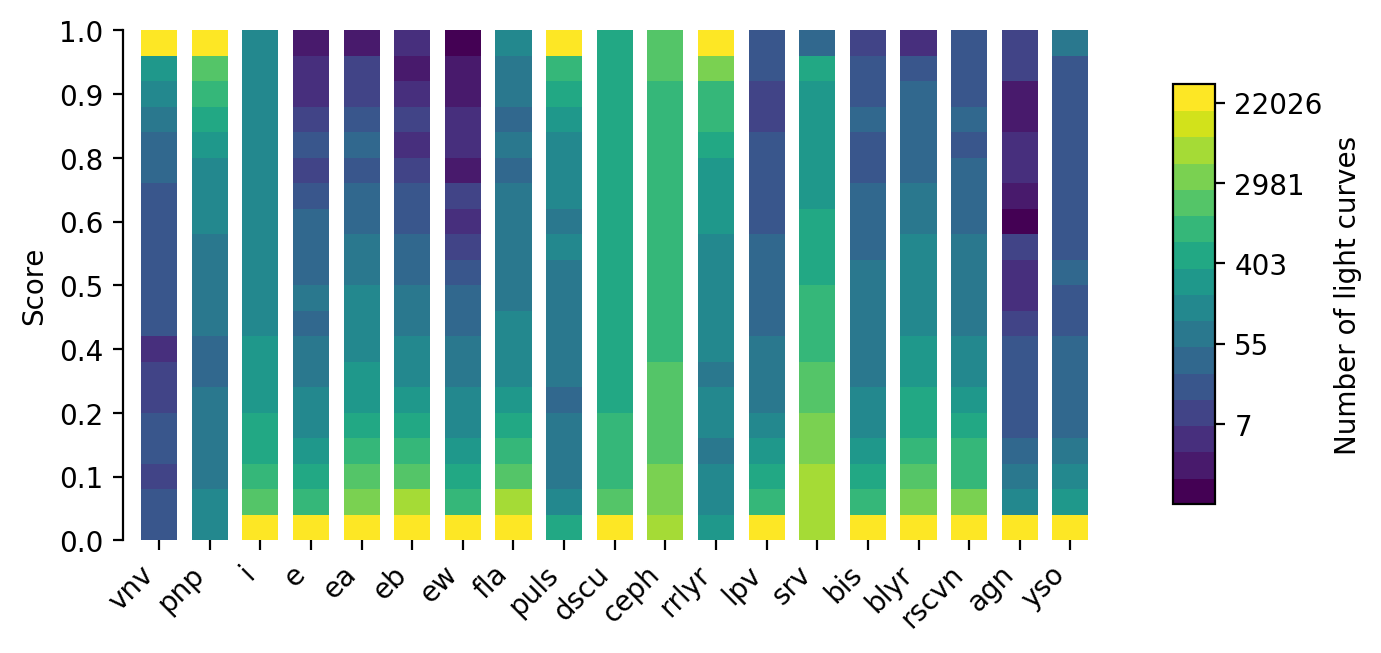}}\quad
  \subfigure[Cepheids]{\includegraphics[width=0.48\textwidth]{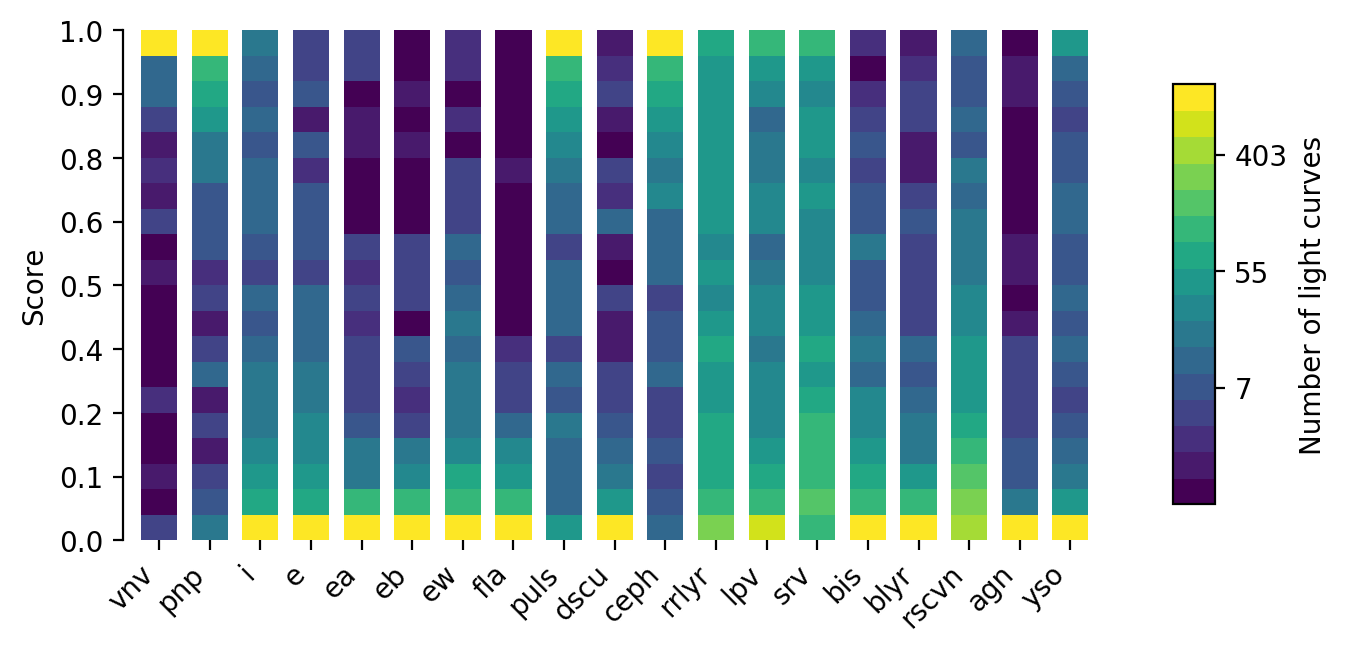}}\quad
\caption{Score distributions color-coded in logarithmic scale of all the DNN classifiers on different object types from the Chen et al. set.}
\label{fig:dnn_on_chen}
\end{figure*}

We also compare our classification scores for the periodic variable classes with the results from \citet{chen2020}. \citet{chen2020} used the ZTF data from Data Release 2 to search for periodic variable stars. They classified the variable stars by comparing the periods and parameters describing the light curve shape. Fig. \ref{fig:dnn_on_chen} shows the classifiers performance assuming the \citet{chen2020} labels as ground truth. The score distributions indicate that our machine learning classifiers mostly agree with the classification by \citet{chen2020}. 

\begin{figure}
     \centering
     \includegraphics[width=\columnwidth]{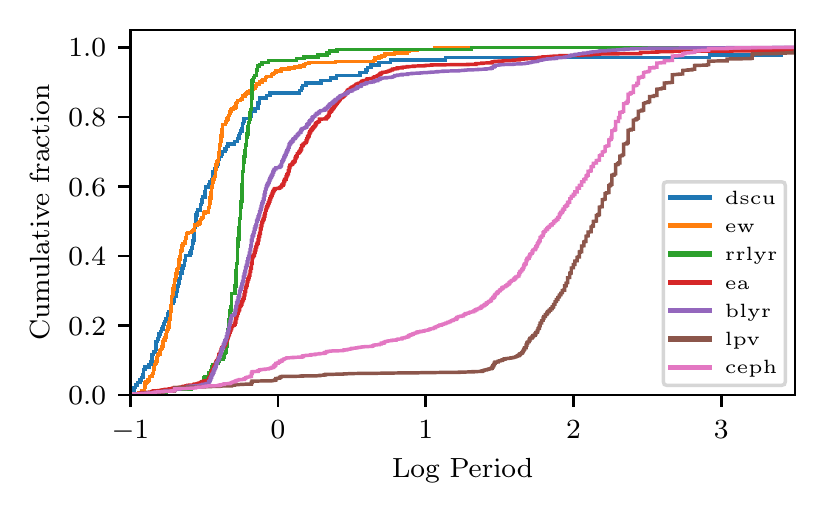}
     \caption{The cumulative period distributions for different classes. The samples have been selected by selecting on the score(variable)$>0.9$ and score([class])$>0.9$ which are \textit{not} in the training-set. The distribution are generally what can be expected for each class. A few percent of systems do have period which are either very short or very long.}
     \label{fig:periodhist}
\end{figure}

As a final sanity check, we inspected how the classifications are distributed. To do this, we selected the most confidently classified objects (we used score(variable)$>0.9$ and score([class])$>0.9$), and plotted them in different feature spaces. First of all, we inspected how the periods are distributed as shown in Fig. \ref{fig:periodhist}. This shows that the periods are generally as expected for the different classes. Only $\approx 5\%$ percent of the periods do not seem to match what is expected for their respective classes. We also inspected the distribution as a function of the number of epochs in the light curves. This shows that the classifiers are not confident for objects with fewer than 100 epochs, but this varies by classifier. We finally inspected the spatial distribution, and it does not show any anomalies.

\section{Discussion}
\subsection{Example usages and real-life performance}
The classifier performance on the test, gold, and external high-purity sets indicate the precision (and to a much lesser extent the recall) of the models, however, say little about the ``real-life'' performance when evaluated on the full corpus of ZTF light curve data. To explore the performance in a production setting, we ran our classifiers on all light curves in the 20 test fields.

\subsubsection{RR Lyrae}
RR Lyrae pulsators are a well-defined class of pulsating stars that are relatively easy to identify. As a typical example, we query classification results to obtain a clean set of RR Lyrae. As criteria we use are: $\mathrm{score(variable)} > 0.9$ and $\mathrm{score(RR Lyrae)} > 0.9$. A total of 2102 out of 34 million light curves pass these criteria. Because objects can have multiple light curves, this corresponds to a total of 1199 astrophysical objects. A visual inspection of the light curves shows that 1073 objects (89\%) are RR Lyrae variables. False positives include a few Delta Scuti stars (22) and Cepheid variables (21) which have the same light curve shape (and are closely related to the RR Lyrae pulsators), but have different pulsation periods. Other false positives are mostly high amplitude irregular variables. 

This analysis shows that the classifier works on `real-life' data. As expected, the real-life performance is slightly lower than the test performance. It also indicates that there is some confusion with irregular variables, which be solved by adding more examples of the latter to the training set.

\subsubsection{Finding YSOs}
Young stellar objects exhibit variability on a wide range of timescales, from hours to months, 
that may be periodic or quasi-periodic when associated with stellar rotation, or aperiodic/irregular when
related to accretion from a circumstellar disk onto the central star, which is a more stochastic process.
Previous attempts to find and classify YSOs using machine learning techniques \citet{} 
have not been particularly successful, having both low completeness and low reliability.

As an example of a challenging classification task, we inspect the sample of high probability YSO's in light curves of the 20 test fields. We select all light curves with $\mathrm{score(YSO)} > 0.9$. A visual inspection shows that approximately 26\% of the classified YSO's can be confirmed as bonafide young variables. Contaminants included AGN/ QSO/ Seyfert classes, which have similar aperiodic variability to YSOs (both object categories are often described as a damped random walk), as well as  pulsating AGB (e.g., Mira and SRVs), post-AGB (e.g., RV Tau)
and other types of LPVs, with which YSOs also share some features. YSOs can also have flaring behavior similar to the CV class, though contamination from this category was $<0.5$\%.

Early tests on all ZTF data, specifically, around the Galactic plane and Gould's Belt regions, the techniques described here show great potential for discovering large numbers of new, previously unappreciated YSOs.

\subsection{Comparison with ZTF alert brokers}
\edit1{
ZTF alert brokers, e.g. ANTARES \citep{Saha2014}, ALeRCE \citep{sanchez2020}, Lasair \citep{Smith2019}, and FINK \citep{2020MNRAS.tmp.3384M}, use ZTF alerts to identify and classify objects which exhibit variability in the ZTF data, a goal similar to that of this project. The approach and focus is different however.
The alert brokers (currently) only use the ZTF alert data, which are generated by 5 standard deviation sources on \textit{difference} images, and lack any information on lower amplitude variability. 
The aim of this work is specifically to identify and classify \textit{all} stars, including low amplitude variables. We therefore use PSF photometry of \textit{all persistent} point-sources in the ZTF science images to classify them all. Therefore, we also use different processing methods (most importantly period finding).}

\subsection{Deficiencies of the classifiers and improvements}
While the classifiers are working very well, we have identified a few deficiencies. 
First, the classification performance drops off for objects which are fainter than 20th magnitude. We expect the classification performance to decrease with magnitude simply because of the lower precision in the light curves. However, inspecting light curves of faint, misclassified variables shows that a human (and thus the machine learning algorithm) is able to easily classify these light curves. Inspection of the training set shows that there is a relative lack of faint variable objects in the training sample (see Fig. \ref{fig:training_set}). With our active learning framework, we are able to remedy this by labeling a set of faint objects with a high variability score.

A second issue is a large number of misclassifications of irregular variables and ``bogus'' objects. In building the training sample, we have focused mostly on identifying periodic variable stars since they are easy to identify. Analyses of the classification results show that many irregular variables that are not classified correctly and also that many seemingly high amplitude variables turn out to be ``bogus'' (internal reflections in the ZTF telescope). We expect to solve this issue automatically while using the classifier; as objects are misclassified, we will encounter them while using the classifiers. They will be added to the training sample, and the next iteration of classifiers will learn to better classify similar false positives.

\subsection{Meta-classification}
In this work, we have run DNN and XGBoost independently. Each of them works very differently and yet it is heartening to see their consistent performance with high precision and recall for almost all classes. The small number of misclassifications are of two types: (a) outliers -- these will be misclassified by both types of classifiers, and (b) objects with a subset of properties not quite captured by the classifier -- these will likely be different for the two classifier types. By combining the classifications from the two classifier types we can obtain even purer samples. The misclassifications - or more specifically their deviant properties - will provide an additional facet to the active learning training regime we have employed here. That will be our next goal as we bootstrap from the sources classified in this work.

\section{Summary and future work}
In this paper, we have established the framework and infrastructure for the machine learning classification of ZTF light curves. In future work, we will use the framework construct the ZTF variable object catalog which provides light curve features and classifications for all ZTF light curves. The catalog will allow astronomers to efficiently search the ZTF light curves for objects of interests. In addition to the catalog, we will also make the training set available for users who wish to run their own classifiers (e.g. Alert brokers). The variable catalog and training set will be updated periodically to incorporate improvements in the classification.

The classification performance will be improved in a number of ways. First of all, as ZTF keeps on accumulating data, both the time baseline and the number of epochs will increase. This will both improve the classification of longer-timescale phenomena, but also allow for the detection of more subtle variability (e.g. the detection of low-amplitude periodic variables or narrow features like eclipses). 

As astronomers are using the classifications and visually inspect the light curves, they will continue to label data. This will further improve the machine learning classifiers by correcting misclassifications and adding them to the training sample. In addition, exploration of the data by using a combination of light curve features and phenomenological classes (e.g. periodic variables that do not fall into any of the known ontological classes), allows us to identify rarer classes and add them to the classification scheme. 

Future work also includes testing of improved and different machine learning methods. Neural networks and XGBoost are currently state of the art, but new machine learning methods are being developed at a rapid pace. The currently implemented methods will serve as a baseline benchmark to test novel methods. For example; recurrent neural networks can be used to classify variable-length time series directly without the need for features. In addition, unsupervised machine learning methods can be applied to find anomalous light curves.

\acknowledgments
We thank the referee for useful and constructive feedback on the manuscript.

Based on observations obtained with the Samuel Oschin Telescope 48-inch and the 60-inch Telescope at the Palomar Observatory as part of the Zwicky Transient Facility project. ZTF is supported by the National Science Foundation under Grant No.  AST-1440341 and a collaboration including Caltech, IPAC, the Weizmann Institute for Science, the Oskar Klein Center at Stockholm University, the University of Maryland, the University of Washington (UW), Deutsches Elektronen-Synchrotron and Humboldt University, Los Alamos National Laboratories, the TANGO Consortium of Taiwan, the University of Wisconsin at Milwaukee, and Lawrence Berkeley National Laboratories. Operations are conducted by Caltech Optical Observatories, IPAC, and UW.

DAD acknowledges support from the Heising-Simons Foundation under Grant No. 12540303.
AAM acknowledges support from the NSF grant OAC-1640818.
MWC acknowledges support from the National Science Foundation with grant number PHY-2010970.
The authors acknowledge support from Google Cloud.

The authors acknowledge the Minnesota Supercomputing Institute\footnote{\url{http://www.msi.umn.edu}} (MSI) at the University of Minnesota for providing resources that contributed to the research results reported within this paper under project ``Identification of Variable Objects in the Zwicky Transient Facility.''
This research used resources of the National Energy Research Scientific Computing Center (NERSC), a U.S. Department of Energy Office of Science User Facility operated under Contract No. DE-AC02-05CH11231 under project ``Towards a complete catalog of variable sources to support efficient searches for compact binary mergers and their products.'' This work used the Extreme Science and Engineering Discovery Environment (XSEDE), which is supported by National Science Foundation grant number ACI-1548562. This work used the Extreme Science and Engineering Discovery Environment (XSEDE) COMET at SDSU through allocation AST200016.

\vspace{5mm}
\facilities{ZTF}


\software{astropy \citep{2018AJ....156..123A},
          keras \citep{chollet2015keras},
          keras-tuner \citep{omalley2019kerastuner},
          kowalski \citep{DuMa2019},
          matplotlib \citep{Hunter:2007},
          numpy \citep{2011CSE....13b..22V},
          pandas \citep{reback2020pandas},
          tensorflow \citep{tensorflow2015-whitepaper},
          xgboost \citep{chen2016}
          }



\clearpage 
\appendix

\section{Features}

\begin{table*}
\begin{center}
 \begin{tabular}{l l l l} 
 \hline
 \# & feature name & description and reference &  \\ 
 \hline\hline
1 & period & best period in days   & \\
2 & significance & significance of the period  & \\
3 & n &  number of epochs in the light curve   & \\
4 & median & median magnitude    & \\
5 & wmean  & weighted mean magnitude    & \\
6 & wstd & weighted standard deviation   & \\
7 & chi2red & reduced $\chi^2$ value after subtracting the mean   & \\
8 & roms & robust mean statistic \citep{rose2007} &  \\
9 & norm peak to peak amp & normalised peak-to-peak amplitude \citep{sokolovsky2009} & \\
10 & norm excess var & normalised excess variance \citep{nandra1997} & \\
11 & MAD & median absolute deviation & \\
12 & IQR & the interquartile range & \\
13 & f60 & the inter-60\% range & \\
14 & f70 & the inter-70\% range & \\
15 & f80 & the inter-80\% range & \\
16 & f90 & the inter-90\% range & \\
17 & skew & the skewness (2nd moment) & \\
18 & smallkurt & the kurtosis (3rd moment) & \\
19 & inv vonneumannratio & the inverse Von Neumann ratio \citep{neumann1941,neumann1942}  & \\
20 & welch i & the Welch I statistic \citep{welch1993} &   \\
21 & stetson j & the Stetson J statistic \citep{stetson1994} &   \\
22 & stetson k & the Stetson L statistic \citep{stetson1994}  &   \\
23 & ad & Anderson Darling test \citep{stephens1974} &   \\
24 & sw & Shapiro Wilk test \citep{shapiro1965} &   \\
25 & f1 power & $\dfrac{\chi_0^2-\chi^2}{\chi_0^2}$ of the fit &   \\
26 & f1 BIC & difference in BIC value \citep{schwarz1978} &   \\
27 & f1 s & slope &   \\
28 & f1 c & constant &  \\
29 & f1 amp & amplitude of the fundamental period &   \\
30 & f1 phi0 & phase of the fundamental period &   \\
31 & f1 relamp1 & relative amplitude of 1st harmonic &  \\
32 & f1 relphi1 & relative phase of 1st harmonic &   \\
33 & f1 relamp2 & relative amplitude of 2nd harmonic&   \\
34 & f1 relphi2 & relative phase of 2nd harmonic &   \\
35 & f1 relamp3 & relative amplitude of 3rd harmonic &   \\
36 & f1 relphi3 & relative phase of 3rd harmonic &   \\
37 & f1 relamp4 & relative amplitude of 4th harmonic &   \\
38 & f1 relphi4 & relative phase of 4th harmonic &   \\
39 & n ztf alerts & number of alerts within 2\arcsec\ \citep{duev2019} &   \\
40 & mean ztf alert braai & the mean `real-bogus' score of the alerts \citep{duev2019}  & \\
41 & dmdt & a 26 by 26 histogram of all dm-dt pairs \citep{mahabal2017} &   \\
\hline
\end{tabular}
\end{center}
\caption{ZTF features we calculated for each of the light curves. The `f1' features are parameters from a fit on the phasefolded light curves. All features are used by the phenomenological classifiers (barring n, the number of points in a light curve). XGBoost excluded \textit{dmdt} as well. See \citealt{CoBu2020} for more information.}
\label{tab:ZTFfeatures}
\end{table*}

 
\begin{table*}
\begin{center}
 \begin{tabular}{l l l l} 
 \hline
 \# & feature name & description and reference &   \\ 
 \hline\hline
1 & AllWISE  w1mpro & W1 magnitude  & \\
2 & AllWISE  w1sigmpro & W1 magnitude uncertainty  & \\
3 & AllWISE  w2mpro &  W2 magnitude  & \\
4 & AllWISE  w2sigmpro & W2 magnitude uncertainty  & \\
5 & AllWISE  w3mpro &  W3 magnitude  & \\
6 & AllWISE  w3sigmpro & W3 magnitude uncertainty  & \\
7 & AllWISE  w4mpro &  W4 magnitude  & \\
8 & AllWISE  w4sigmpro & W4 magnitude uncertainty  & \\
9 & AllWISE  ph qual & Photometric quality flag  & \\
10 & Gaia DR2  phot g mean mag & G-band mean magnitude  & \\
11 & Gaia DR2  phot bp mean mag & BP-band mean magnitude  & \\
12 & Gaia DR2  phot rp mean mag & RP-band mean magnitude & \\
13 & Gaia DR2  parallax & absolute stellar parallax  & \\
14 & Gaia DR2  parallax error & parallax uncertainty  & \\
15 & Gaia DR2  pmra & proper motion in right ascension   & \\
16 & Gaia DR2  pmra error & standard error of proper motion in right & \\
17 & Gaia DR2  pmdec & proper motion in declination   & \\
18 & Gaia DR2  pmdec error & standard error of proper motion in declination   & \\
19 & Gaia DR2  astrometric excess noise & Excess noise of the source  & \\
20 & Gaia DR2  phot bp rp excess factor & BP/RP excess factor  & \\
21 & PS1 DR1  gMeanPSFMag & Mean PSF AB magnitude from $g$ filter   & \\
22 & PS1 DR1  gMeanPSFMagErr & Error in the magnitude from $g$ filter    & \\
23 & PS1 DR1  rMeanPSFMag & Mean PSF AB magnitude from $r$ filter   & \\
24 & PS1 DR1  rMeanPSFMagErr &  Error in the magnitude from $r$ filter   & \\
25 & PS1 DR1  iMeanPSFMag & Mean PSF AB magnitude from $i$ filter   & \\
26 & PS1 DR1  iMeanPSFMagErr & Error in the magnitude from $i$ filter    & \\
27 & PS1 DR1  zMeanPSFMag & Mean PSF AB magnitude from $z$ filter   & \\
28 & PS1 DR1  zMeanPSFMagErr & Error in the magnitude from $z$ filter    & \\
29 & PS1 DR1  yMeanPSFMag & Mean PSF AB magnitude from $y$ filter   & \\
30 & PS1 DR1  yMeanPSFMagErr & Error in the magnitude from $y$ filter  & \\
31 & PS1 DR1  qualityFlag & binary flag denoting if real of false positive  & \\
\hline
\end{tabular}
\end{center}
\caption{Features external to ZTF. Barring the quality flags these were used in the ontological classifiers in addition to using the ZTF features.}
\label{tab:Externalfeatures}
\end{table*}



This section shows all the light curve statistics (Table \ref{tab:ZTFfeatures}) and external statistics (Table \ref{tab:Externalfeatures}). The statistics and processing of light curves is discussed in detail in \cite{Coughlin2020}. In this appendix, we present the equations of non-standard statistics or statistics for which multiple definitions exists. We refer readers the appropriate references in the Table in other cases. In the equations, we use $m$ for the magnitudes, $t$ for the observation times, $N$ for the total number of epochs, and $i$ to indicate individual epochs.

\subsection{Amplitude statistics}
We calculate a few simple statistics which are measures of amplitude: MAD (median absolute deviation), the inter-quartile range, and the inter-percentile ranges for 60, 70, 80, and 90 percent.

\begin{equation}
    \textrm{MAD} = \textrm{median} \left( |m_i - m_{\textrm{median}}| \right)
\end{equation}

\begin{equation}
   \textrm{IQR} = \mathrm{percentile}(m_i,75\%)-\mathrm{percentile}(m_i,25\%)
\end{equation}

\begin{equation}
    \textrm{f90} = \mathrm{percentile}(m_i,95\%)-\mathrm{percentile}(m_i,5\%)
\end{equation}

\subsection{Higher order moments}
We calculate the higher order moments with the equations given below.
\begin{equation}
 \textrm{Skewness} = \frac{N}{(N-1) (N-2)} \sum\limits_{i} \frac{\left(m_{\textrm{mean}}-m_i \right)^{3}}{\sigma_i^3}
 \end{equation}
 
 \begin{equation}
 \textrm{Kurtosis} = \frac{N (N+1)}{(N-1) (N-2) (N-3)} \sum\limits_{i} \frac{\left(m_{\textrm{mean}}-m_i \right)^{4}}{\sigma_i^4} - \frac{3 \left(N-1\right)^{2}}{(N-2) (N-3)}
 \end{equation}

\subsection{Von Neumann ratio}
The Von Neumann ratio measures the ratio between the correlated variance and the variance.
\begin{equation}
\eta = \left( \sum\limits_{i} \left(\frac{1}{\Delta t_i} \right)^2 m_{\textrm{var}} \right)^{-1}\sum\limits_{i} \left( \frac{\Delta m_i}{\Delta t_i} \right)^2
\end{equation}
with $\Delta t_i = t_{i+1} - t_{i}$ and $\Delta m_i = m_{i+1} - m_{i}$ \\

\subsection{Welch \& Stetson statistics}
The use the Welch-Stetson I and Stetson J \& K statistics from \citet{stetson1996}


\begin{equation}
\begin{split}
\delta_i & = N/(N-1) (m_i-\textrm{wmean}) \\ 
P_i & = \delta_i \delta_{i+1} \\
J & = \sum sign(P_i) \sqrt{|P_i|} \\
K & = \sum(|\delta_i|)/N/\sqrt{1/N \sum \delta_i^2} \\
\end{split}
\end{equation}

\section{Class description}
\subsection{Phenomenological classes}\label{sec:phenomenological_classes}
\begin{itemize}
    \item \textbf{Variable (vnv)}; a `variable' source is any ZTF source which shows variability in its light curve due to \textit{astrophysical} origin. This excludes variability due to blended photometry, bright nearby stars, or any CCD artifact. In a sense, this step can be regarded as a 'real-bogus' filter \citep{duev2019}.
    \item \textbf{Periodic (pnp)}; the ZTF light curve features astrophysical periodic variability. These are typically pulsators, rotating stars and (eclipsing) binaries. This does not include semi-periodic or quasi-periodic variability. This excludes variability due to a varying background (e.g. due to the moon), or spurious periodic variability due to nearby bright stars or other artifacts. 
    \item \textbf{Flaring (fla)}; any ZTF light curve that shows flares, rapidly rising and fading events. These are mostly cataclysmic variables, some young stellar objects, some AGN.
    \item \textbf{Irregular (i)}; objects which show irregular variability. These are mostly accreting objects, AGN, CVs, and YSOs. 
    \item \textbf{Long timescale (longt)}; any object which shows variability on timescales of 100 days. This can be a steady increase/decrease in luminosity, e.g., AGN and CVs. Long timescale period variable like Miras and the more irregular semi-regular variables are generally included in this category.
    \item \textbf{Eclipsing (e)}; any source which shows eclipses in the light curve. These are predominately eclipsing binaries. Eclipsing planetisimals or planets with large rings would also fall in this category. The subtypes are \textbf{EW} (overcontact binaries), \textbf{EB} (semi-detached binaries), and \textbf{EA} (detached binaries).
    \item \textbf{Bogus}; any light curve that seems variable, but is not due to any astrophysical variability. These are galaxies which can seem to vary due to PSF variations, image artifacts like `ghosts', blended stars, or diffraction spikes. 
\end{itemize}

\subsection{Ontological classes}\label{sec:Ontological_classes}
\begin{itemize}
    \item \textbf{Active Galactic Nuclei (agn)}; extra-galactic objects which tend to vary irregularly. Often show slowly rising or fading light curves, and can also show outbursts in rare cases.
    \item \textbf{Long Period Variables (lpv)}; Long Period Variables are cool giant stars. Nearly all stars of this type show some variability. \textbf{Mira} variables are AGB stars which show very high amplitude ($>2.5$), long period (80-1000 days) variability. \textbf{Semi Regular Variables (\textbf{srv})} show more irregular, and lower amplitude variability than Miras.
    \item \textbf{Pulsator (puls)}; any kind of pulsating star.
    \item \textbf{Cepheid (ceph)}; Cepheids are radially pulsating giant stars in the instability strip. Period range between 1 and 50 days, with extreme examples of 200 days. light curves shapes range from asymmetric with a steep rise and slow decay, to almost sinusoidal light curve shapes.
    \item \textbf{Delta Scuti (dscu)}; Delta Scuti are pulsating A\&F main sequence stars. The pulsation period ranges between 0.03 between 0.3 days, and the amplitude is typically 0.2 mag but can reach up to 0.8mag. Their light curves are asymmetric, with a rapid rise and slow decay. 
    \item \textbf{RR Lyrae (rrlyr)}; RR Lyrae are radial pulsators on the horizontal branch; they are helium core burning and hydrogen shell burning. The pulsation period ranges between 0.2 and 1.0 days. \textbf{RR Lyrae ab} are pulsating in their fundamental mode and show amplitudes of up to 1 magnitude and have asymmetric light curves with a steep rising phase. \textbf{RR Lyrae c} are first overtone pulsators. They have maximum magnitudes of 0.5, and show more sinusoidal light curves.  \textbf{RR Lyrae d} pulsate at two periods ('d' stand for `double').   \textbf{RR Lyrae Blazkho} are RR Lyrae which show evolution in their light curve shape, known as the Blazkho effect. 
    
    \item \textbf{Binary star (bis)}; any object which is a binary star.    
    \item \textbf{RS CVn (rscvn)}; a binary star in which at least one of the components has large stellar spots. The light curve shape is sinusoidal with periods of a few hours to 14 days. The shape of the light curve changes over timescales of months to years.
    \item \textbf{Beta Lyrae (blyr)}; Binary systems were one of the components has evolved into a subgiant or giant star and is filling it's Roche lobe transferring mass in a disk. The light curve of these systems is of type 'EB'. The period ranges between 0.3 days and 200 days. Systems with a period of $>$100 days contains a supergiant.
    \item \textbf{Young Stellar Objects (yso)}; pre-main-sequence stars. They typically have an accretion disk and dust around them. This results in a light curve which shows irregular behavior, sometimes with outbursts or dust obscuration events. 

\end{itemize}

\section{Overview of the 20 test-fields.}
In Table \ref{tab:testfields} we show basic properties of the test-fields we used. They were selected as pairs and chosen to make sure that they represent different Galactic latitudes and ZTF coverage.


\begin{table}
\centering
\begin{tabular}{ l | rrrrrrr }
FieldID & ra & dec & l & b & $\#g$ & $\#r$ & $\#i$ \\
\hline
296 & 15.7910 & -17.05 & 141.274 & -79.1454 & 126 & 165 & 23 \\
297 & 22.8729 & -17.05 & 168.678 & -75.7386 & 108 & 156 & 26 \\
423 & 192.250 & -2.65 & 303 & 59.95 & 85 & 126 & \\
424 & 199.340 & -2.65 & 316.926 & 59.1829 & 89 & 115 &  \\
487 & 281.193 & 4.55 & 36.5577 & 3.0284 & 230 & 598 & 52 \\
488 & 288.119 & 4.55 & 39.7486 & -3.0953 & 221 & 584 & 52 \\
562 & 88.373 & 18.95 & 189.844 & -2.9775 & 245 & 595 & 16 \\
563 & 95.578 & 18.95 & 193.164 & 2.9745 & 266 & 1047 & 14  \\
682 & 266.856 & 33.35 & 58.6098 & 26.7346 & 822 & 790 & 79 \\
683 & 274.709 & 33.35 & 60.8553 & 20.5066 & 773 & 1013 & 61 \\
699 & 45.975 & 40.55 & 148.921 & -15.1819 & 220 & 430 & 6 \\
700 & 54.523 & 40.55 & 154.444 & -11.5358 & 241 & 430 & 6 \\
717 & 200.867 & 40.55 & 96.8087 & 75.0527 & 613 & 654 & 166 \\
718 & 209.484 & 40.55 & 80.2113 & 70.6464 & 636 & 683 & 167 \\
777 & 49.315 & 54.95 & 143.261 & -1.7248 & 324 & 709 & 4 \\
778 & 60.116 & 54.95 & 148.223 & 1.9883 & 300 & 577 & 4 \\
841 & 145.714 & 69.42 & 142.515 & 40.204 & 368 & 354 & 31 \\
842 & 162.857 & 69.35 & 137.094 & 44.7025 & 390 & 359 & 31\\
852 & 334.286 & 69.35 & 110.25 & 10.5906 & 262 & 292 & \\
853 & 351.429 & 69.35 & 115.729 & 7.9313 & 232 & 269 & \\
\end{tabular}
\caption{The ZTF-fields IDs and statistics for the fields we have used for testing our procedures while developing the pipeline}
\label{tab:testfields}
\end{table}

\bibliographystyle{aasjournal}
\bibliography{references,zotero} 

\end{document}